\documentclass[a4paper,11pt]{article}
\pdfoutput=1 % if your are submitting a pdflatex (i.e. if you have
\usepackage{jheppub} % for details on the use of the package, please
\usepackage{float}
\restylefloat{table}
\usepackage[T1]{fontenc} % if needed
\usepackage{relsize}
\usepackage{slashed}
\usepackage{ytableau}
\usepackage{amsthm}
\usepackage[utf8]{inputenc}
\usepackage{cleveref}
\crefformat{section}{\S#2#1#3} % see manual of cleveref, section 8.2.1
\crefformat{subsection}{\S#2#1#3}
\crefformat{subsubsection}{\S#2#1#3}

\usepackage[force]{feynmp-auto}

\usepackage{braket}
\usepackage{filecontents}

\def \12x12 {$\frac{1}{2}\otimes \frac{1}{2}$ }
\def \0x1 {$0\otimes1$ }
\newcommand{\sSlm}[3]{{}_{#1} S_{#2 #3}}

\font\ec=ecrm0800 at 11pt
\def\th{\hbox{\ec\char'336}}
\def\edth{\hbox{\ec\char'360}}

\title{Scattering in Black Hole Backgrounds and Higher-Spin Amplitudes: Part II  }

\author[a,b,c]{Yilber Fabian Bautista,}\emailAdd{yilber-fabian.bautista-chivata@ipht.fr}

\author[d,e]{Alfredo Guevara,}\emailAdd{aguevaragonzalez@fas.harvard.edu}

\author[f,g]{Chris Kavanagh}\emailAdd{chris.kavanagh1@ucd.ie }

\author[e,g,h]{and Justin Vines}\emailAdd{justin.vines@aei.mpg.de }

\affiliation[a]{Perimeter Institute for Theoretical Physics, Waterloo, ON N2L 2Y5, Canada}
\affiliation[b]{Department of Physics   and  Astronomy, York University, Toronto, Ontario, M3J 1P3, Canada}
\affiliation[c]{Institut de Physique Théorique, CEA, Université Paris–Saclay,
F–91191 Gif-sur-Yvette, France}
\affiliation[d]{Center for the Fundamental Laws of Nature,  Society of Fellows
\& Black Hole Initiative, Harvard University, Cambridge, MA 02138, USA}

\affiliation[e]{Kavli Institute for Theoretical Physics, University of California, Santa Barbara, CA, 93106-4030, USA}
\affiliation[f]{School of Mathematics and Statistics,
University College Dublin, Belfield, Dublin 4, Ireland}
\affiliation[g]{Max Planck Institute for Gravitational Physics (Albert Einstein Institute), Am  M\"{u}hlenberg 1, Potsdam 14476, Germany}
\affiliation[h]{Mani L. Bhaumik Institute for Theoretical Physics, Department of Physics and Astronomy,
UCLA, Los Angeles, USA}

\abstract{ We continue to investigate correspondences between, on the one hand, scattering amplitudes for massive higher-spin particles and gravitons in appropriate quantum-to-classical limits, and on the other hand, classical gravitational interactions of spinning black holes according to general relativity. We first construct an ansatz for a gravitational Compton amplitude, at tree level, constrained only by locality, crossing symmetry, unitarity and consistency with the linearized-Kerr 3-point amplitude, to all orders in the black hole's spin.  We then explore the extent to which a unique classical Compton amplitude can be identified by comparing  with the results of the classical process of scattering long-wavelength gravitational waves off an exact Kerr black hole, determined by appropriate solutions of the Teukolsky equation.
Up to fourth order in spin, we find complete agreement with a previously conjectured exponential form of the tree-level Compton amplitude.  At higher orders, we extract tree-level contributions from the Teukolsky amplitude by an analytic continuation from a physical ($a/GM<1$) to a particle-like ($a/GM>1$) regime. Up to the sixth order in spin, we identify a unique \textit{conservative} part of the amplitude which is insensitive both to the choice of boundary conditions at the black hole horizon and to branch choices in the analytic continuation. The remainder of the amplitude is determined modulo an overall sign from a branch choice, with the sign flipping under exchanging purely ingoing and purely outgoing boundary conditions at the horizon.
 Along the way, we make contact with novel applications of massive spinor-helicity variables pertaining to their relation to EFT operators and (spinning) partial amplitudes.
}

\begin{document} 
\maketitle
\flushbottom

\section{Introduction}

Progress in the study of relativistic celestial mechanics, especially of systems of black holes (BHs) in classical general relativity (GR), continues to be enriched by novel applications of the principles of on-shell relativistic quantum scattering amplitudes. In turn, there is arguably great untapped potential for further advances in the understanding of amplitudes arising from work in classical GR, particularly in exploiting the nonperturbative results provided by the exact stationary black hole (Schwarzschild and Kerr) spacetimes, and their linear perturbations.

In unitarity-based approaches to  perturbative amplitudes for the two-spinning-BH problem, key roles are played by the classical limits of the 
%minimal coupling (in the sense of \cite{Arkani-Hamed:2017jhn}),  
 tree-level, $n$-point amplitudes $A_n$:  these involve one massive spinning matter line (representing a single BH, both incoming and outgoing) meeting $n-2$ gravitons. The simplest building blocks, the 3-point amplitude $A_3$ and the Compton amplitude $A_4$, %--- the 3-point amplitude: BH in, meets graviton, BH out; and the Compton amplitude: BH and graviton in, graviton and BH out ---
completely determine the classical dynamics of spinning binary BH systems up to  the second post-Minkowskian (PM) order \cite{Vaidya:2014kza,Bjerrum-Bohr:2018xdl,Cheung:2018wkq,Kosower:2018adc,Guevara:2017csg,Guevara:2018wpp,Chung:2018kqs,Bautista:2019tdr,Maybee:2019jus,Chung:2019duq,Bern:2020buy,Chung:2020rrz,Aoude:2020ygw,Chen:2021kxt,Kosmopoulos:2021zoq,Bautista:2021inx,Chen:2022yxw,Menezes:2022tcs,FebresCordero:2022jts,Aoude:2022trd,Aoude:2022thd,Alessio:2022kwv}.

The appropriate classical 3-point amplitude has been argued to be fixed to all orders in the BH spin at tree level \cite{Chung:2018kqs,Guevara:2018wpp,Arkani-Hamed:2019ymq,Bern:2020buy} by its correspondence with a linearized Kerr solution \cite{Levi:2015msa,Vines:2016qwa,Harte:2016vwo,Vines:2017hyw}.  Remarkably, the same 3-point amplitude $A_3$ arose in \cite{Arkani-Hamed:2017jhn}, independently of any consideration of BHs, (in the large-$S$ limit for massive spin-$S$ particles meeting a graviton) as the unique amplitude with a well behaved high-energy limit smoothly connecting to the corresponding massless amplitude, the latter fixed from  kinematics considerations alone \cite{Benincasa:2007xk}.  This link has sparked significant interest in exploring the space of amplitudes, chiefly 4-point Compton amplitudes $A_4$, which may effectively describe classical gravitational interactions of spinning black holes, in search of certain strongly constraining properties of the amplitudes which may single out the appropriate black-hole solutions.

The tree-level gravitational Compton amplitude $A_4$ has been argued to be fixed for spinning massive particles  up to  $S=3/2$, by only unitarity and consistency with the established 3-point amplitude \cite{Arkani-Hamed:2017jhn,Chung:2018kqs}. Its classical limit has been well studied \cite{Guevara:2018wpp,Aoude:2020onz,Chung:2018kqs,Bautista:2019tdr}, providing an amplitude that is consistent with the linearized BH metric. The classical Compton amplitude has also been computed from effective worldline theory, finding agreement for the case of BHs up to cubic order in spin \cite{Saketh:2022wap}. For spin $S=2$ matter, contact deformations are allowed, for both the quantum \cite{Chung:2018kqs} and classical amplitude \cite{Aoude:2022trd}, then forbidding uniquely fixing the Compton amplitude from unitarity considerations only. 
For higher spin orders  ($S>2$), the conjectural extrapolation of the BCFW Compton amplitude is well known to suffer from unphysical singularities, which can nevertheless  be cured  by contact deformations. In reference \cite{Chiodaroli:2021eug}, a well-behaved  $S=5/2$ quantum  Compton amplitude was computed by requiring consistent  higher-spin current factorization. More recently, the authors of \cite{Bern:2022kto,Aoude:2022trd,Aoude:2022thd} have shown  the remaining freedom of the classical higher-spin Compton amplitude to be strongly constrained by additional  properties, concerning the high-energy limit and a conjectured \textit{spin-shift-symmetry}, which are well-motivated by lower spin orders, but which have not yet been fully shown to correspond to BHs in GR.  Additional constraints from higher spin gauge symmetry have been considered in \cite{Cangemi:2022bew}.

In order to fully establish a  correspondence between the spinning $A_n$ amplitudes --- or equivalent specifications of couplings in effective worldline theories in the PM \cite{Liu:2021zxr,Jakobsen:2022fcj,Goldberger:2017ogt,Jakobsen:2021lvp,Jakobsen:2021zvh,Jakobsen:2022psy,Jakobsen:2022zsx} and post-Newtonian (PN) \cite{Blanchet:2013haa,Porto:2016pyg,Levi:2018nxp,Levi:2015ixa,Levi:2016ofk, Levi:2020kvb, Antonelli:2020aeb,Levi:2020uwu, Antonelli:2020ybz, Kim:2021rfj,Levi:2019kgk,Levi:2020lfn,Cho:2022syn,Cho:2021mqw,Kim:2022pou,Kim:2022bwv,Levi:2022dqm,Levi:2022rrq} approaches --- and classical spinning BHs,  matching computations that go beyond the stationary limit need to be done. A first step  has been taken  along these lines by the authors of \cite{Siemonsen:2019dsu}, where, up to the third  order in the BH's spin, results for the 2PM aligned spin  scattering angle --- computed in \cite{Guevara:2018wpp} from the minimal coupling $A_3$ and $A_4$ ---  were shown to be consistent with the general-relativistic “self-force” calculations of the linear perturbations of a Kerr spacetime sourced by a small orbiting body \cite{Kavanagh:2016idg,Bini:2018ylh}; in particular, the gravitational observables,  Detweiler redshift \cite{Detweiler:2008ft}, and the circular orbit precession frequency  were considered for the matching computation in \cite{Siemonsen:2019dsu}. However, the aforementioned observables depend on particular pieces of $A_4$,  and their matching does not fully characterize the amplitudes for BHs. 

In this paper we remove these obstructions by studying the classical Compton amplitude in the context of  the scattering of a gravitational plane wave off a Kerr BH.  This classical process, like the aforementioned self-force calculations, 
 is approached  using the tools of black hole perturbation theory (BHPT) \cite{Pound:2021qin}. Considering only one massive object allows us to remove the difficulties associated with the inclusion of the second body, therefore providing cleaner connections between the $A_n$  amplitudes and their classical counterparts.

The problem of the scattering of plane waves off  BHs  has been approached from a variety of  perspectives \cite{PhysRevD.16.237,Westervelt:1971pm,Doran:2001ag,Dolan:2007ut,Matzner1968,Chrzanowski:1976jb, PhysRevD.13.775,Guadagnini:2008ha,Westervelt:1971pm}. In a modern on-shell language, in Part I of this work \cite{Bautista:2021wfy}, we have argued that the low-energy content of   plane  wave perturbations of the Kerr BH,  is entirely captured by the classical limit of  4-point amplitudes for the scattering of a massive particle of  infinite spin  and  massless particles of helicity $h<2$, interchanging gravitons. For scalar wave perturbations ($h=0$), we showed that the Born amplitude (presented to all orders in the BH's spin) exactly matches the \textit{conservative} part of the  Teukolsky result up to second order in the BH's spin. (We consider as usual a BH of mass $M$ and spin $J$, and define the ring-radius $a=J/M$, counting orders in spin as orders in $a$.) Starting at order $a^3$, the Born amplitude receives contact deformations that  can be \textit{uniquely}  fixed from the  Teukolsky solution;  we showed explicit results up to order $a^5$. 
Our extraction of the scattering amplitudes from a BHPT computation relies on analytically extending   results for $a^\star:=a/GM<1$  (where the tools of BHPT are well defined) to the $a^\star\gg1$ region, where the BH ceases to have a horizon, exposing a naked singularity. However, this analytic extension has to be performed in a careful manner since horizon dissipation effects (present at order $a^1$! for the scalar case), get mixed with \textit{conservative} contributions by the analytic continuation. In Part I we presented a prescription for extracting the \textit{conservative} contribution to the amplitude, by removing dissipative effects before the analytic continuation. As a further complication, the BHPT results are discontinuous at $a^\star=1$, then, a choice of branch for doing the extension through this discontinuity in the complex $a^\star$ domain, needs to be taken. This choice of  branch is however irrelevant for the extraction of  the \textit{conservative} contribution, whereas, as we will show in this paper for the  $h=2$ case,  the dissipative parts (kept if desired) come with an additional $\eta=\pm$ sign,   fixed by  the branch choice. 

This paper is organized as follows. In \cref{sec:GWscatt_1} we study the problem of the scattering of a gravitational wave off a Kerr BH from a  covariant Compton amplitude, warming up through quadratic order in spin. This is obtained through  the  spin-multipole double copy 
of \cite{Bautista:2019tdr,Bautista:2021inx}, which we  review and extend in \cref{eq:spin_dc}. In this section, we set some notation and clarify previously raised tensions  between the BHPT 
\cite{Dolan:2008kf} and a  Feynman diagramatic \cite{Guadagnini:2008ha,Barbieri:2005kp}  approach to the gravitational wave scattering problem, at linear order in the BH's spin. In order to provide an amplitude extendible to all orders in spin, in \cref{compton-spin} we switch to the spinor-helicity construction of the  $A_n$ amplitudes.  In \cref{eq:spin_exponential} we present spin-exponentiated formulas for the quantum amplitudes valid to all orders in spin for the $n=3$ case, and up to  $S=2$ for $n=4$. For the latter case, we further study the factorization properties of the amplitudes. 
In \cref{sec:classical_limit} we show how to extract the classical contributions from these amplitudes, and argue that by imposing that the same classical amplitude is extracted for quantum amplitudes written in   the chiral or anti-chiral spinor bases, the classical amplitude is constrained by a classical version of the QFT crossing symmetry (see also \cref{ap:chiral} and \cref{ap:bose}). In \cref{Sec:General Compton Amplitudes} we present an ansatz for the (classical) Compton amplitude for generic spins, based on physical  assumptions: unitarity and locality together with classical crossing symmetry. In \cref{sec:shift_symmetry} we briefly discuss on the recently conjectured \textit{spin-shift-symmetry} \cite{Bern:2022kto,Aoude:2022trd,Aoude:2022thd}, and show 
 our ansatz can be made consistent with previous analyses.  In \cref{sec:matchTA} we summarize the  results from   matching to a full BHPT computation. We provide explicitly 
 a \textit{unique, conservative} Compton amplitude up to order $a^6$, and furthermore identify dissipative contributions, encapsulated by special operators in the Compton ansatz, proportional to $|a|\omega$.  The resulting amplitudes are not invariant under the \textit{spin-shift-symmetry}. 
In \cref{Sec:PWScattering} (see also \cref{PW-App}) we discuss the classical (BHPT) computation, showing explicit examples to illustrate the partial wave decomposition, and the analytic extension to the $a^\star\gg1$ region, together with the identification of the \textit{conservative} and \textit{dissipative} contributions. In \cref{sec:mathcing_comtpon} we discuss the matching of the BHPT solutions to the Compton ansatz  (see also \cref{sec:spinningspheroidalharmonics}, where this matching is done by using \textit{massive spinor-helicity} variables to represent the harmonics). We conclude with a discussion in \cref{sec:discussion}.

\section{Gravitational wave scattering from the $a^2$-covariant amplitude  }\label{sec:GWscatt_1}

It is illustrative to study the problem of scattering gravitational waves off the Kerr metric by comparing known results in GR to the tree-level gravitational Compton amplitude $A_4$, written in vector notation. In this section we will do so up to quadratic order in the BH spin, which will be useful to layout the general approach to the classical limit, while establishing notation/conventions.

Amplitudes written covariantly in terms of tensors have the characteristic of encoding  graviton helicity states  ($h=\pm2$), in  symmetric and traceless polarization tensors $\epsilon^{(\pm)\mu\nu}$, which in 4-dimensions can be written as  the outer  product of two photon-polarization vectors, $\epsilon^{(\pm)\mu\nu}=\epsilon^{(\pm)\mu}\epsilon^{(\pm)\nu}$. Up to quadratic order in the spin parameter ($\mathcal{O}(a^2)$), the classical piece of the   gravitational Compton amplitude was given  in  a spin multipole expansion in \cite{Bautista:2019tdr,Bautista:2021inx}, and whose \textit{double-copy} derivation was further expanded in \cite{Bautista:2022ewu} (see  \cref{eq:spin_dc} for a review). In this section, we will make use of such amplitude and connect it to the classical process of the scattering of a gravitational plane wave off the Kerr BH. Before going into the main computation, we introduce some notation and kinematics conventions. 

\subsection{Kinematic conventions and classical variables}\label{sec:kinematics}
\begin{figure}
\begin {center}
\includegraphics[width=7truecm]{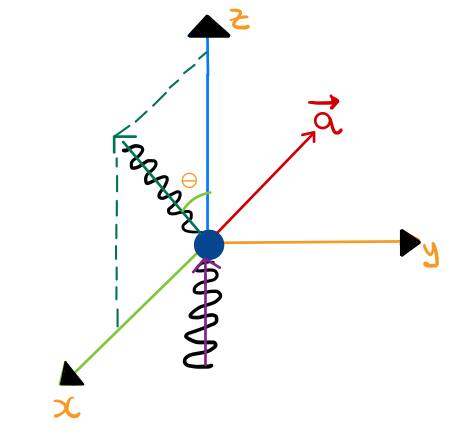} 
\end{center}
\caption{ Gravitational wave scattering setup.  An incoming plane wave  traveling along the $z$-axis hits a  Kerr BH whose spin has an arbitrary orientation $\vec{a}$, and is initially  at rest; the wave gets scattered with outgoing  momentum lying in the $x-z$ plane.    
}
\label{fig:ampl}
\end{figure}

\begin{figure}
\begin {center}
\includegraphics[width=4truecm]{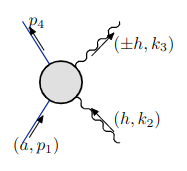} 
\end{center}
\caption{ Kinematic conventions for the scattering of a gravitational wave  ($h=\pm2$) off the  Kerr BH.  
}
\label{eq:A4-figure}
\end{figure}

Let us start by introducing the kinematic variables we will use when writing  amplitudes explicitly in terms of the scattering angle. The classical process we have in mind is the scattering of a gravitational plane wave off a Kerr BH which is initially at rest and has   a spin  vector arbitrarily orientated, $\vec{a}=(a_x,a_y,a_z)$, see \cref{fig:ampl}.  
From a QFT setup,  this classical process can be treated by thinking of the classical compact object as a massive, spinning point particle, whereas the waves can be thought of as the incoming/outgoing massless particles of helicity $\pm2$ composing the gravitational Compton amplitude (See \cref{eq:A4-figure}). 
We  label the    incoming/outgoing massless  momenta  entering in the gravitational Compton amplitude as $k_2/k_3$, whereas the spinning massive particle   has initial/final momenta $p_1/p_4$, mass $M$, and  spin parameter $a^\mu=s^\mu/M$, the later corresponding  then to the radius of the ring singularity of the Kerr BH,  in a classical setup, whereas the massless particles are  identified as gravitational waves of a given helicity (circular polarization states).

The classical observable relevant to this scattering process is the differential cross-section, which we choose   to evaluate  in the reference frame for which the initial BH is at rest, and the scattering process happens in the $x-z$ plane.  This then leads us to choose the explicit representation for the kinematic variables  (see \cite{Bautista:2021wfy} for more details).

\begin{equation}
\label{eq:kinematics}
    \begin{split}
    p_{1}^{\mu} & =(M,0,0,0),\\
k_{2}^{\mu} & =\hbar \omega(1,0,0,1),\\
k_{3}^{\mu} & =\frac{\hbar \omega(1,\sin\theta,0,\cos\theta)}{1+2\frac{\hbar \omega}{M}\sin^{2}(\theta/2)},\\
p_{4}^{\mu} & =p_{1}^{\mu}+k_{2}^{\mu}-k_{3}^{\mu}\,.  
    \end{split}
\end{equation}
 Here  the form of the energy of the outgoing wave, $k_3$, is fixed by the on-shell condition for the outgoing massive momentum. The independent kinematic invariants  can be put in terms of  these variables as follows:
\begin{equation}\label{eq:sandtclassical}
    \begin{split}
     s & =(p_1+k_2)^2=M^{2}\left(1+2\frac{\hbar \omega}{M}\right)\,,\\
t & =(k_3-k_2)^2=-\frac{4\hbar^2 \omega^{2}\sin^{2}\left(\theta/2\right)}{1+2\frac{\hbar \omega}{M}\sin^{2}(\theta/2)}\,.   
    \end{split}
\end{equation}
When writing the Compton amplitude in a spinor-helicity basis, it will also be useful to introduce the optical parameter
\begin{equation}\label{eq:xidef}
    \xi^{-1}:=-\frac{M^2 t}{(s-M^2)(u-M^2)}=- \sin^2(\theta/2)+\mathcal{O}(\hbar) \,.
\end{equation}
The optical parameter  will help us to infer the behavior of the classical amplitude for certain limit values of the scattering angle $\theta$. In particular, the so-called eikonal limit is obtained as $\xi\to \infty$.

 Let us also write the 
4-dimensional graviton polarization tensors  from the product of photon polarization vectors in a given  circular polarization states. In terms of the scattering angle $\theta$, we choose the explicit basis for the incoming states
\begin{equation}\label{eq:photon polarization2}
    \begin{split}
        \epsilon_{2}^{+}= & \frac{1}{\sqrt{2}}(0,1, i,0) \,,\\
\epsilon_{2}^{-} =& - \frac{1}{\sqrt{2}}(0,1,- i,0)\,. 
    \end{split}
\end{equation}
Analogously,  for the outgoing states we have
\begin{equation}\label{eq:photon polarization1}
    \begin{split}
        \epsilon_{3}^{+}= & =\frac{1}{\sqrt{2}}(0,\cos\theta, i,-\sin\theta)\, ,\\
\epsilon_{3}^{-}  =&
=- \frac{1}{\sqrt{2}}(0,\cos\theta,- i,-\sin\theta)\,. 
    \end{split}
\end{equation}

% \begin{figure}
%  \begin{align*}\nonumber
%  \begin{fmffile}{a3a4}\hspace{0.5cm}
%   \hspace{0.3cm}\,\,
%   \,\,
%   \hspace{0.06cm}
%   \parbox{22pt}{
%   \begin{fmfgraph*}(70,70)
%     \fmfleft{i2,i1}
%     \fmfright{o2,o1}
%     \fmftop{t}
%     \fmf{phantom,tension=0.2}{i1,v1,i2}
%     \fmf{phantom}{o1,v2,o2}
%     \fmf{phantom,tension=0.3}{v1,v2}
%     \fmffreeze
%     \fmf{plain,width=0.7
%     ,foreground=(0.035,,0.168,,0.623)}{g,i1}
%     \fmf{plain,width=0.7,tension=2.8}{g,v1}
%     \fmf{plain,width=0.7,foreground=(0.035,,0.168,,0.623)}{i2,v1}
%     \fmf{photon,width=0.7,tension=2.8}{o1,v1}
%     \fmf{photon,width=0.7}{o2,v1}
%     \fmfv{decor.shape=circle,decor.filled=12,decor.size=.35w}{v1}
%     \marrow{eb}{right}{rt}{$(\pm h,k_3)$}{v1,o1}
%     \marrow{ed}{right}{rt}{$(h,k_2)$}{o2,v1}
%     \marrow{ea}{up}{top}{$p_4$}{g,i1}
%     \marrow{ec}{down}{bot}{$(a,p_1)$}{i2,v1}
%   \end{fmfgraph*}}
%  \hspace{3.5cm}
%    \end{fmffile}
% \end{align*}
% \caption{ Kinematic conventions for the scattering of a gravitational wave  ($h=\pm2$) off the  Kerr BH.  }
% \label{eq:A4-figure}
% \end{figure}

It was deeply discussed in  \cite{Bautista:2021wfy} how taking  the classical limit of the QFT amplitude reduces to take $\hbar\omega/M\ll1$, and the spin of the black hole $a/\hbar\gg1$, while keeping the product $\omega a$ fixed. With this in mind, the  classical differential cross-section can be computed from the formula 
\begin{equation}
\frac{d\langle\sigma\rangle}{d\Omega}=\sum_{h^\prime} \frac{|\langle A_{4}(h\rightarrow  h^\prime)\rangle|^{2}}{64\pi^{2}M^{2}}\,,\label{eq:classical cross section}
\end{equation}
where the angle brackets, $\langle\rangle$, indicate only the classical piece of the  amplitude is to be included in the previous formula.

It will also be convenient to introduce a basis of spinor helicity variables for the massless legs of Figure \ref{eq:A4-figure} in terms of the scattering angle $\theta$. We  fix the little group freedom in such a way that  using the massless spinors
\begin{equation}\label{eq:spinhel}
    \begin{split}
      |\mu\rangle&= \left(e^{-i\phi/2} \cos\theta/2, e^{i\phi/2} \sin\theta/2 \right),\,\,\, |\lambda\rangle = \sqrt{2\omega}\left(-e^{-i\phi/2} \sin\theta/2, e^{i\phi/2} \cos\theta/2 \right),\\
      [\mu| &=\left(e^{i\phi/2} \cos\theta/2, e^{-i\phi/2} \sin\theta/2 \right),\,\,\, [\lambda| = \sqrt{2\omega}\left(-e^{i\phi/2} \sin\theta/2, e^{-i\phi/2} \cos\theta/2 \right)\,,
    \end{split}
\end{equation}
to construct  massless momentum and polarization matrices, via
\begin{eqnarray}\label{eq:vectospin}
     k_{\alpha \dot{\alpha}} = |\lambda\rangle_{\alpha} [\lambda|_{\dot \alpha}\, &,\\
       \epsilon^{+}_{\alpha \dot{\alpha}} = \sqrt{2} \frac{|\mu \rangle_{\alpha} [\lambda|_{\dot \alpha}}{\langle \lambda \mu\rangle }\, &,\epsilon^{-}_{\alpha \dot{\alpha}}=  -\sqrt{2}\frac{|\lambda \rangle_{\alpha} [\mu|_{\dot \alpha} }{[\mu \lambda]}   \,,
\end{eqnarray}
the states associated to $k_3,k_2$ are recovered by setting $\phi=0$ for the former, and additionally $\theta=0$ for the latter. These spinor-helicity variables will play a crucial role in \cref{sec:spinningspheroidalharmonics}  as we will use them to construct \textit{spinning spherical harmonics}.

\subsection{Classical scattering cross section}\label{sec:classical_cross_section}
With previous notation at hand, as well as the ground rules for the extraction of the classical limit of the required  QFT amplitudes, we are ready to tackle the low energy regime for the problem of  scattering of a gravitational wave off the Kerr BH and at  low orders in the  spinning multipole expansion, from a modern on-shell perspective. Our aim is to compute the  classical differential cross section \eqref{eq:classical cross section}. For that, we need the   classical piece of the Compton amplitude. In   \cite{Bautista:2022ewu} (see also \cref{eq:spin_dc}), it was showed that up to order $a^2$, and on the support of the Spin supplementary condition (SSC) $p_{\cdot}a=0$, this amplitude takes the following  form
\begin{equation}
    \langle A_4 ^{\text{GR}}\rangle = \frac{\kappa^2}{8}\frac{\langle\omega^{(0)}\rangle}{k_2{\cdot}k_3 (p_1{\cdot}k_2)^2}\left[\langle \omega^{(0)} \rangle + \langle \omega^{(1)\mu\nu}\rangle\epsilon_{\mu\nu\rho\sigma}p_1^{\rho}a^{\sigma} + \langle \omega^{(2)}_{\alpha\beta} \rangle a^{\alpha}a^{\beta}   \right]+ \mathcal{O}(a^3)\,.\label{eq:a4CLASSSPIN2}
\end{equation}
The  multipole  coefficients $\omega^{(i)}$ are given explicitly in \eqref{eq:w0},  \eqref{eq:w1}, and \eqref{eq:quad_class}, and the angular bracket notation    indicates we have to take the classical limit  of the corresponding spin multipole  coefficients. Notice remarkably,  amplitude \eqref{eq:a4CLASSSPIN2} written in this fashion has the  the factorization form $A^{\text{GR}} = K_4 \langle A_0^{\text{QED}}\rangle \times \langle A_s^{\text{QED}}\rangle $, as suggested by the authors  in \cite{Bern:2020buy}. This is a manifestation of the equivalence principle; that is, in \cite{Bautista:2022ewu}, the gravitational amplitude was derived  from the double copy of two spin $\frac{1}{2}$ QED Compton amplitudes;  in the  classical limit, however, this product can be rearranged into the separation  of the scalar piece and the spin pieces.

The next  task  is to use the kinematics 
 (\ref{eq:kinematics} - \ref{eq:photon polarization1}) into \eqref{eq:a4CLASSSPIN2} and show that the  different helicity configurations for the wave scattering process results into the following amplitudes 
\begin{align}
    \langle A_{4}^{++}\rangle &= \frac{\kappa^2 M^2\cos^4(\theta/2)}{4\sin^2(\theta/2)}\big[1+\mathcal{F} (\omega,a,\theta)+\frac{1}{2!}\mathcal{F}(\omega,a,\theta)^2+\mathcal{O}(a^3) \big],\label{eq:App-gr}\,,\\
     \langle A_{4}^{- -}\rangle&=  \left[\langle A_{4}^{++}\rangle^*\right]_{\omega\rightarrow-\omega}\,,\\
    \langle A_{4}^{+ -}\rangle&= \frac{\kappa^2 M^2\sin^4(\theta/2)}{4\sin^2(\theta/2)}\big[1+\mathcal{G}(\omega,a,\theta) +\frac{1}{2!}\mathcal{G}(\omega,a,\theta)^2+\mathcal{O}(a^3)\big]\,,\\
     \langle A_{4}^{- +}\rangle&=  \left[\langle A_{4}^{+ -}\rangle^*\right]_{\omega\rightarrow-\omega}\label{eq:Amp-gr}\,,
\end{align}
where the super-scripts label the helicity of the incoming/outgoing waves, and we have   used 
\begin{eqnarray}
     \mathcal{F}(\omega,a,\theta) &=& -2a_z\omega\sin^2(\theta/2) +a_x\omega\sin\theta-2(a_x-ia_y)\omega\tan(\theta/2)\,,\label{eq:f-function}\\
     \mathcal{G}(\omega,a,\theta) &= &2a_z\omega\sin^2(\theta/2) -a_x\omega\sin\theta \,.
\end{eqnarray}
The structure of these amplitudes is  nothing by the quadratic in spin truncation of the exponential functions $e^{\mathcal{F}(\omega,a,\theta)}$ and $e^{\mathcal{G}(\omega,a,\theta)}$, which appear  naturally from a spinor-helicity rewriting of the different helicity  configuration of the Compton amplitude, as we will discuss in \cref{compton-spin}. Notice extrapolation of the previous formulas up to  $\mathcal{O}(a^4)$ provides perfectly well-defined amplitudes for all values of the scattering angle\footnote{Of course as $\theta\to0$, we hit the usual  t-channel divergence. }. However, extrapolation to $\mathcal{O}(a^5)$, results into the unphysical backwards scattering singularity ($\theta\to\pi$) of $\langle A_4^{++}\rangle$ due to the $\tan{\theta/2}$ term in \eqref{eq:f-function}. Removal of such singularity will be the topic of study of \cref{compton-spin}.

Using the  amplitudes \eqref{eq:App-gr}-\eqref{eq:Amp-gr} into \eqref{eq:classical cross section}, we can compute the unpolarized classical differential cross section  for the scattering of GW off the Kerr BH,  which up to  quadratic order in spin takes the form\footnote{The differential cross section  can be easily extended up to  $\mathcal{O}(a^4)$ as discussed above.}
\begin{equation}\label{eq:sigmaspin2}
\begin{split}
     \frac{d\braket{\sigma}}{d\Omega} = \frac{G^2M^2}{\sin^4(\theta/2)}\Big[\cos^8(\theta/2)\Big(1+2\tilde{\mathcal{F}}+2\tilde{\mathcal{F}}^2\Big) +
     \sin^8(\theta/2)\Big(1+2\mathcal{G}+2 \mathcal{G}^2\Big)\Big]+\mathcal{O}(a^3)\,,
\end{split}
\end{equation}
with $\tilde{\mathcal{F}}=\mathcal{F}\big|_{a_y=0}$. Interestingly, the only spin components contributing to the actual observables are those with non-zero projection on the scattering plane, whereas the off-plane components are just phases of the amplitude.

It is illustrative to consider for instance the differential cross section    for the polar scattering configuration \footnote{In this configuration, the incoming wave has momentum aligned with the direction of the spin of the BH, which we choose to be $a^\mu=(0,0,0,a_z)$. }, let us for simplicity print here the linear in-spin contribution 

\begin{equation}
    \frac{d\braket{\sigma}}{d\Omega}\Big|_{\text{polar}}{=}\frac{G^{2}M^{2}}{\sin^{4}(\theta/2)}\Big[\cos^{8}(\theta/2)\left(1{-}4 a_z\omega\sin^{2}(\theta/2)\right){+}\sin^{8}(\theta/2)\left(1{+} 4 a_z\omega\sin^{2}(\theta/2)\right)\Big]{+}\mathcal{O}(a^2).
\label{eq:cross-section-gravity-kerr}
\end{equation}
This  recoveres the   black hole perturbation theory (BHPT) result derived by  Dolan in \cite{Dolan:2008kf}, from totally different arguments. Our result however disagrees with the Feynman diagrammatic derivation of  Barbieri and Guadagnini provided in   \cite{Guadagnini:2008ha,Barbieri:2005kp}. The solution of the tension raised in \cite{Dolan:2008kf}, at this order in spin from the classical and the Feynman diagrammatic approach  is  solved -- as already announced in \cite{Bautista:2021wfy} -- by  including  all of the Feynman diagrams contributing to the gravitational Compton amplitude, and not just the graviton exchange diagram, as considered  by the authors of \cite{Guadagnini:2008ha,Barbieri:2005kp}. This agreement of the QFT derivation of the classical differential cross section and the BHPT results goes in fact beyond the leading in-spin order or the polar scattering configuration. As we will show below, extrapolation of \eqref{eq:sigmaspin2} to $\mathcal{O}(a^4)$ indeed agrees with the BHPT derivation, and in  \cref{compton-spin} we will expand on the amplitudes computations to arbitrary spin, with a final unique \textit{conservative} answer fixed up to the sixth power in the spin of the BH.

The fact that the Compton amplitude computation recovers the  result from BHPT is non-trivial. In fact, as we will thoroughly expand in \cref{Sec:PWScattering}, but also observed for the scalar case \cite{Bautista:2021wfy}, in order to obtain the differential cross section from BHPT one needs to deal with very complex  intermediate  steps, starting from  the expansion of the scattering amplitudes into infinite sums of harmonics, to the  change of  basis from the spin weighted spheroidal harmonics into the spin weighted spherical harmonics, in which case, at a given order in spin the infinite sum truncates \textit{only in the polar scattering case}. Then, the QFT computation efficiently resumes these infinite series of harmonics in a very simple expression. On the other hand, we have obtained the result for the generic orientation of the spin of the black hole, which to the best of our knowledge  has not been obtained from  BHPT. The reason for this is that setting the off-axis problem makes the computation much more complicated since now the infinite sums have two running indices, as well as the complicated additions of expanding gravitational plane waves in a basis of spheroidal harmonics as discussed in \cref{PW-App}.

\section{Compton Amplitude for Arbitrary Spins}\label{compton-spin}

We have seen how the Compton amplitude matches previous results obtained in \cite{Dolan:2008kf} for GW scattering off a Kerr BH,  at linear order in spin in the   polar scattering configuration, hence solving the disagreement with previous Feynman diagrammatic approaches \cite{Guadagnini:2008ha,Barbieri:2005kp}. The matching suggests that the GW cross-section studied in \cite{Dolan:2008kf} carries much of the information of the classical limit of the Compton amplitude. In \cref{Sec:PWScattering} we will study the classical scattering amplitude for a non-polar incident wave by solving the $h=-2$ Teukolsky equation and argue that in the low energy regime and up to a phase (unimportant for the physical observable \eqref{eq:classical cross section}),  the Compton amplitude  indeed captures the same information obtained from the classical computation. However, as the non-polar amplitude will be given in terms of an (infinite) partial wave expansion, additional input from QFT is needed in order to extract the relevant information, as we extensively discussed  in \cite{Bautista:2021wfy} for the scalar case. 

In this section, we will write the most general form of the tree-level Compton amplitude for any spin $S$ as allowed by locality and unitarity, under the classical limit. This in turn will fix the unphysical ($\theta\to\pi$) singularity for $S\ge5/2$ of the naive BCFW exponential form of the amplitude \cite{Guevara:2018wpp,Aoude:2020onz}. 
At each order in spin, we will see there are only  a few effective operators that survive the limit, hence only a few free coefficients will have to be fixed from the Teukolsky computation. The resulting amplitude will effectively resum the full partial wave expansion as we shall show in \cref{Sec:PWScattering}.

In the previous section, we studied the quadratic in-spin amplitude in vector notation as obtained from the spin-multipole double copy. However, this double-copy prescription is not easy to implement for  higher-spinning scenarios. Therefore, 
to study general arbitrary spins it is more convenient to work with the massive spinor-helicity variables of \cite{Arkani-Hamed:2017jhn}.\footnote{These were first implemented in this context in \cite{Guevara:2017csg}, see also more recent developments in \cite{Guevara:2018wpp,Chung:2018kqs,Chung:2019duq,Arkani-Hamed:2019ymq}}
We start by reformulating the classical limit in terms of massive spinors, more precisely in terms of $\textrm{SU}(2)$ operators. For $S\leq 2$ our construction will match the classical limit obtained by Aoude et al for three and four-point amplitudes \cite{Aoude:2020onz} employing a Heavy Particle Effective Theory (HPET). For $S>2$, after studying the classical limit we will directly construct the ansatz at the strict $\hbar \to 0$ limit, although an analytic continuation to $\hbar \neq 0$ is trivial to find.

To start, consider again massive particles with momenta $p_ 1 ,p_4$.  Let $\sigma^{\mu}$ ($\tilde{\sigma}^{\mu}$) be the chiral (antichiral) set of Pauli matrices, so that we have, for instance
\begin{equation}
    P_1:=p_1^{\mu}\sigma_{\mu} = |1^{a_1}\rangle [1^{a_2}| \epsilon_{a_1 a_2}\,,\qquad\,  \tilde{P}_1:=p_1^{\mu}\tilde{\sigma_{\mu}} = |1^{a_1}]\langle 1^{a_2}| \epsilon_{a_1 a_2} \,,\end{equation}
where $a_1,a_2=1,2$, which transform under the massive little group $\textrm{so}(3)\sim \textrm{su}(2)$. Indices are raised and lowered via the 2-dimensional antisymmetric tensor $\epsilon_{ab}$. The 2-component spinors $|1^a\rangle$ are normalized by $-[1^11^2]{=}\langle 1^1 1^2\rangle {=}M$ and hence solve the Dirac equation in momentum space:
\begin{equation}\label{eq:p1map}
    \tilde{P}_1|1^{a}\rangle = M |1^a]\,,\quad  P_1|1^{a}] = M |1^a\rangle \,.
\end{equation}
Spin-$S$ polarization states can be represented as totally-symmetric tensor products:
\begin{align}
    |\varepsilon_1 \rangle =& \frac{1}{M^S} |1^{(a_1}\rangle \otimes \ldots \otimes |1^{a_{2S})}\rangle  \label{basi1} \,, \\
     |\varepsilon_1 ] =& \frac{1}{M^S} |1^{(a_1}] \otimes \ldots \otimes |1^{a_{2S})}] \,,
\end{align}
which are two different choices for a basis of $2S+1$ states. They can be mapped to each other using the operator \eqref{eq:p1map}. In these bases, we can define the `scattering matrix' $A_n^{\textrm{chir}}$ and its conjugate operator $A^{\textrm{antichir.}}_n$ by
\begin{equation}\label{eq:chiral-to-antichiral}
    A_n^{h=2,S}= \langle \varepsilon_n | A^{\textrm{chir.}}_n | \varepsilon_1 \rangle =  [ \varepsilon_n | A^{\textrm{antichir.}}_n | \varepsilon_1 ]\,,
\end{equation}
where $p_n$ is the final massive momentum. 

\subsection{Exponentiated Amplitude for $n=3,4$}\label{eq:spin_exponential}

Before studying the classical limit we perform a multipole decomposition of the amplitudes for $n=3,4$ \cite{Guevara:2018wpp}. The (quantum, tree-level) minimal coupling amplitudes were there given in terms of the angular momentum operator $\mathbb{ J}^{\mu \nu}$ as

\begin{equation}\label{ampli}
        A^{S}_{3}=A^{0}_3 \times \langle \varepsilon_3 | \exp\left(\frac{F_{2\mu \nu}\mathbb{ J}^{\mu\nu}}{2i\epsilon_2\cdot p_1}\right) | \varepsilon_1 \rangle\,, A^{S}_{4}=A^{0}_4 \times \langle \varepsilon_4 | \exp\left(\frac{F_{2,\mu \nu}\mathbb{ J}^{\mu\nu}}{2i \epsilon_2 \cdot p_1}\right) | \varepsilon_1 \rangle \,.
\end{equation}
Recall that the graviton polarization tensors are given by $\epsilon_i^{\mu \nu}=\epsilon_i^\mu \epsilon_i^\nu$ and we have defined the momentum-space field strength tensors $F_i^{\mu \nu}=2k_i^{[\mu}\epsilon_i^{\nu]}$.  Momentum conservation reads, respectively,
\begin{equation}
    \begin{split}
    p_3&= p_1+k_2  \\
p_4 &= p_1 + k_2 + k_3  \,.
    \end{split}\label{34kins}
\end{equation}
In both cases we assume the graviton associated to $k_2$ to have negative helicity, and for $n=4$ the graviton $k_3$ has positive helicity.\footnote{In \cref{sec:kinematics} the $k_3$ graviton had opposite momentum as compared to conventions here. To connect to previous section, we simply take $k_3\to-k_3$ here,  which also flips its helicity from positive to negative. We stick to such prescription here as it will be more convenient to express crossing symmetry.} (The same-helicity configuration will be also discussed momentarily) In this case it is convenient to fix the gauge, in spinor-helicity variables, as \cite{Guevara:2018wpp}

\begin{equation}\label{eq:chkgauge}
    \epsilon_2=\frac{\sqrt{2} |3]\langle 2|}{[32]} \propto \tilde{\epsilon}_3 = \frac{\sqrt{2} |3]\langle 2|}{\langle32\rangle }\,.
\end{equation}
The operator $\mathbb{ J}_{\mu\nu}$ in \eqref{ampli} the spin-$S$ Lorentz generator. In this case, we will realize it as a fully quantum operator acting linearly on the representation $|\varepsilon\rangle$. The exponential series truncates at order $2S$ in the expansion of the exponential. The pole $\epsilon_2\cdot p_1$ will cancel in the cases treated here from the scalar amplitude $A_n^0$. As anticipated, this effectively restricts $S\leq 2$ in the Compton amplitude $A_4$.

Before we proceed with the computation, it is illustrative to make some comments on minimal coupling. Recall \textit{minimal coupling} in the sense of \cite{Arkani-Hamed:2017jhn}, is defined by requiring scattering amplitudes involving massive particles  to have a well-defined  high energy limit (free of mass-divergences). This means, massive spin-$S$ spinors  turn into  their massless helicity $h=S$ analogs under such a limit, which in turn implies massive spin-$S$ amplitudes  reduce to  their massless helicity$-S$ counterparts. While this can be shown to be true  at 3-points for generic $S$, where the massless 3-point amplitude is  fully fixed from little group arguments  \cite{Benincasa:2007xk}, at 4-points the statement is  satisfied only up to $S=2$. For instance, in  
the high energy limit,  $A_4^{S=2}$ reduces  to the 4-graviton scattering amplitude in Einstein Gravity. 
For $S>2$, amplitude $A_4^S$ in \eqref{ampli} suffers from mass divergences \cite{Chung:2018kqs}, which make the amplitude have an ill-defined high energy limit, therefore corresponding to interactions  involving non-elementary particles. These divergences,  manifest themselves as the  unphysical poles, $(\epsilon_2{\cdot}p_1)^l$ with $l>2$, appearing in the amplitude when the exponential  is expanded at $\mathcal{O}(\mathbb{ J}^{l})$. These unphysical poles  can be removed by adding a series of contact terms in the amplitude with arbitrary coefficients. Since no massless 4-point amplitude can be uniquely fixed from little group arguments for $S>2$, the strict high energy limit cannot be used to fix  these free coefficients.\footnote{Although a relaxed version thereof has been recently addressed in \cite{Chiodaroli:2021eug}.} Therefore, alternative ways   need to be found in order to fully fix the $S>2$ amplitude at 4-pts.  The matching of the classical 4-point amplitude to the  GW scattering process in General Relativity, as shown  in the previous section, suggests that in the classical limit those free coefficients can be uniquely fixed from solutions of the Teukolsky equations. In this work, we will follow this intuition and find a unique conservative\footnote{In following sections we will also keep track of  dissipative effects whose matching to a Compton ansatz is unique up to a sign dependent of the prescription taken to analytically extend the  BHPT solutions into the point particle regime of the Kerr BH.  }   Compton amplitude up to sixth order in spin. 
As we will see in detail, at $S=2$ the 4-point amplitude in \eqref{ampli} admits non-minimal coupling deformations that preserve unitarity properties of the Compton amplitude  \cite{Chung:2018kqs}, therefore corresponding to contact operators entering in the amplitude (See \cref{Sec:General Compton Amplitudes}). These $S=2$ contact operators can also be fixed from the Teukolsky computation.

Let us  now introduce a four-velocity vector $u^\mu$ together with a generic mass scale $m$. They will be mapped to the four-velocity and rest frame mass of the classical object, respectively. However, the identification with the kinematic momenta in the Compton amplitude is ambiguous, some choices are  $u=\frac{p_1}{m}, \frac{p_4}{m}, \frac{p}{m}$ (with $m=M$ in the former cases, and $m^2=p^2,p=\frac{p_1+p_4}{2}$ in the latter), which will all coincide after we take the classical limit. Now, in order to parametrize the degrees of freedom associated with spin in four dimensions, we introduce the Pauli-Lubanski operator

\begin{equation}\label{spinvec}
    \mathbb{W}^{\mu} :=\frac{1}{2m} \epsilon^{\mu \nu \rho \sigma } u_{\nu} \mathbb{ J}_{\rho \sigma}
\end{equation}

In the classical limit, this will play the role of the spin vector $a^\mu\equiv \langle\mathbb{W}^\mu\rangle$, introduced in the previous section. Using spinor-helicity variables we can   find  additional exact quantum relations between operators. For this, note that in  \eqref{ampli} the field strength $F_2^{\mu\nu}$ is self-dual since the graviton $k_2$ has negative helicity. Consequently, the generator $\mathbb{J}_{\mu \nu}$ is also self-dual and it is associated with the chiral basis \eqref{basi1}, i.e. $\mathbb{J}^{\mu \nu}=\frac{i}{2} \epsilon^{\mu \nu \rho \sigma}\mathbb{J}_{\rho \sigma}$. \footnote{More precisely, we have \cite{Guevara:2018wpp}

\begin{equation}
     \langle \varepsilon_4 | \exp\left(\frac{F_{2,\mu \nu}\mathbb{J}^{\mu\nu}}{2i \epsilon_2 \cdot p_1}\right) | \varepsilon_1 \rangle =  [ \varepsilon_4 | \exp\left(\frac{F_{3,\mu \nu}\tilde{\mathbb{J}}^{\mu\nu}}{2i \epsilon_3 \cdot p_1}\right) | \varepsilon_1 ]\,, \nonumber
\end{equation}

i.e. using the negative helicity graviton also changes the chirality of the Lorentz generator.
} We use this property to rewrite the exponents of \eqref{ampli} in terms of the Pauli-Lubanski operator \eqref{spinvec} as follows. Following \cite{Bautista:2019evw}, for a given 4-velocity $u^{\mu}$ we decompose the full Lorentz generator $\mathbb{J}^{\mu \nu}$ into a spin and a boost operator:

\begin{equation}\label{eq:BandS}
\mathbb{B}^\mu:=\mathbb{J}^{\mu \nu} u_\nu\, ,\qquad \mathbb{S}^{\mu \nu}:=\mathbb{J}^{\mu \nu} - 2 u^{[\mu} \mathbb{B}^{\nu]} \,.
\end{equation}
 One can easily check that $u_{\mu}\mathbb{S}^{\mu \nu}=0$, hence $\mathbb{S}^{\mu\nu}$ generates little group transformations on states $|\varepsilon\rangle$ and shall be related to the Pauli-Lubanski operator $\mathbb{W}^{\mu}$. Indeed, from \eqref{spinvec} one easily finds
\begin{equation}\label{aprop}
     \mathbb{W}^{\mu} =\frac{1}{2m} \epsilon^{\mu \nu \rho \sigma } u_{\nu} \mathbb{S}_{\rho \sigma} \Leftrightarrow \mathbb{S}^{\mu \nu} = - m \epsilon^{\mu \nu \rho \sigma} u_{\rho} \mathbb{W}_{\sigma} \,.
\end{equation}
Furthermore, due to the self-dual condition on $\mathbb{J}^{\mu \nu}$, it turns out that the boost and spin parts are indeed related. From \eqref{spinvec} and \eqref{eq:BandS} we find:

\begin{equation}\label{bprop}
    \mathbb{B}^\mu = i m \mathbb{W}^\mu \,.
\end{equation}
We can now decompose the exponent of \eqref{ampli}. We proceed for both $n=3,4$ at the same time, introducing the generic field strength $F_{\mu \nu}=2 k_{[\mu} \epsilon_{\nu]}$. Using \eqref{aprop} and \eqref{bprop} we have
\begin{eqnarray}
F_{\mu \nu} \mathbb{J}^{\mu \nu} &=& F_{\mu \nu} \mathbb{S}^{\mu \nu} + 2 u_\mu F^{\mu \nu} \mathbb{B}_\nu \nonumber \\
&=& -m \, \epsilon^{\mu \nu \rho \sigma} F_{\mu \nu} u_\rho \mathbb{W}_\sigma + 2i m\, u_\mu F^{\mu \nu} \mathbb{W}_\nu \,.
\end{eqnarray}
Regarding $F_{\mu \nu}$ as self dual, which follows from the contraction with $\mathbb{J}^{\mu \nu}$ on the LHS, we finally get

\begin{equation}\label{eq:Fj}
  F_{\mu \nu} \mathbb{J}^{\mu \nu}   = \pm 4im\,  u_\mu F^{\mu \nu} \mathbb{W}_\nu \,.
\end{equation}
The $\pm$ sign accounts for self-duality or anti self-duality of the Lorentz generator $\mathbb{J}_{\mu \nu}$, or equivalently, the helicity associated to $F_{\mu \nu}$. We remark that the classical limit has not yet been applied as we have explicitly used the operator notation. Note that the LHS of \eqref{eq:Fj} does not depend on the four-vector $u^\mu$, which we are free to choose. In any case, for $u=\frac{p_1}{M}, \frac{p_4}{M}, \frac{p}{m}$ we can now rewrite \eqref{ampli} as

\begin{equation}\label{ampli2}
    A^{S}_{3}=A^{0}_3 \times \langle \varepsilon_3 | \exp\left(2\frac{u\cdot F_2 \cdot \mathbb{W}}{u\cdot \epsilon_2}\right) | \varepsilon_1 \rangle\,, A^{S}_{4}=A^{0}_4 \times \langle \varepsilon_4| \exp\left(2\frac{u\cdot F_2 \cdot \mathbb{W}}{u\cdot \epsilon_2}\right) | \varepsilon_1 \rangle 
\end{equation}
For $n=3$, the on-shell condition for the outgoing massive momenta imposes $u\cdot k_2=0$. This automatically implies that the pole $u\cdot \epsilon_2$ cancels and we have

\begin{equation}\label{3ptex}
    A^{S}_{3}=A^{0}_3 \times \langle \varepsilon_3 | e^{-2 k_2 \cdot \mathbb{W}}  | \varepsilon_1 \rangle \,.
\end{equation}
For $n=4$, the pole does not cancel in the exponential, as $u\cdot k_2\neq 0$ generically. Since the prefactor $A_4^0$ contains a term $(u\cdot \epsilon_2)^4$, the form \eqref{ampli2} is valid only up to quartic order in the expansion of the exponential, i.e. up to spin $S=2$. A convenient way to encode the unphysical pole is introducing the vector \cite{Aoude:2020onz}

\begin{equation}\label{eq:wdef}
    w^\mu := \frac{u\cdot k_2}{u\cdot \epsilon_2} \epsilon^\mu_2\,,
\end{equation}
so that
\begin{equation}\label{eq:4ptexpdef}
    A^{S}_{4}=A^{0}_4 \times \langle \varepsilon_4 | e^{2(w\cdot \mathbb{W} - k_2\cdot \mathbb{W})} | \varepsilon_1 \rangle \,.
\end{equation}

Let us comment briefly on the  factorization properties in this formula. Note that s-channel factorization follows from the fact that $w\cdot \mathbb{W} \to 0$, which together with the usual argument for $A_4^0$ yields
\begin{equation}
       A^{S}_{4} \rightarrow \frac{1}{2p_1\cdot k_2} A_3^{0,L} A_3^{0,R}  \langle \varepsilon_4 | \varepsilon_I\rangle \langle \varepsilon^I | e^{ -2 k_2\cdot \mathbb{W}} | \varepsilon_1 \rangle \
\end{equation}
where the polarizations $\varepsilon^I$ correspond to the internal particle of spin S.\footnote{The usual projection of the propagator into spin S states is not needed since the operator $e^{ -2 k_2\cdot \mathbb{W}}$ preserves the spin S of the external particle. In other words we can replace $| \varepsilon_I\rangle \langle \varepsilon^I| \leftrightarrow \mathbb{I}$.} We can identify the factor $A_3^{0,L}\langle \varepsilon_4 | \varepsilon_I\rangle = A_3^{S,L}$ as the three-point amplitude for negative helicity \cite{Guevara:2018wpp} thus making factorization explicit. The u-channel factorization follows analogously from crossing symmetry. Now, for the $t=\langle 23\rangle [23]$ channel pole we need the following observation: From \eqref{eq:chkgauge} and \eqref{eq:wdef} it follows that
    \begin{align}\label{eq:obst1}
        w^\mu \to k_2^\mu \,\quad& \textrm{as} \quad \langle 23\rangle \to 0 \nonumber \\
         w^\mu \to -k_3^\mu \,\quad& \textrm{as} \quad [ 23] \to 0 \,,
    \end{align}
Thus we find 
\begin{equation}
    A_4^S \to \frac{1}{t} \begin{cases} 
A_3^0 A_3^G\langle \varepsilon_4 | e^{-2(k_3+ k_2)\cdot \mathbb{W}} | \varepsilon_1 \rangle\quad \textrm{as}\,\, [23]\to 0 ,\\
A_3^0 \bar{A}_3^G \langle \varepsilon_4 | \varepsilon_1 \rangle \quad \textrm{as} \qquad \langle 23\rangle\to 0 \,,
\end{cases}
\end{equation}
where $A_3^G (\bar{A}_3^G) $ corresponds to the MHV (anti-MHV) graviton amplitude. We see that the spin factor deforms as expected for each chirality. Thus we have shown that the expression \eqref{eq:4ptexpdef} is consistent with factorization (unitarity) to all orders in the spin $\mathbb{W}$. However, the 4-pt itself is non-local starting at $S=2$ and hence needs to be corrected using contact terms. In the following we will deal with those in the classical limit, where the above factorization can also be realized explicitly.

\subsubsection*{Boosted basis}

The polarization states $|\varepsilon_1\rangle,\langle\varepsilon_4|$ are associated with initial and final momentum, $p_1,p_4$ respectively. It will be convenient to rewrite them as associated to the 4-velocity $u^\mu$ \cite{Bautista:2019evw}. For instance, taking $u=\frac{p_1}{M}$, we can write
\begin{equation}
    p_4= e^{i\mu M  (k_2+k_3)\cdot \mathbb{B} } p_1=  e^{-\mu M^2 (k_2+k_3)\cdot \mathbb{W} } p_1\,,
\end{equation}
Here $\mu$ is a scalar which explicit expression we do not need, but which is given explicitly in \cite{Guevara:2019fsj}. The analogous formula holds for $n=3$; in this case three-particle kinematics yields $\mu M^2 =1$, hence
\begin{equation}
    p_3=  e^{- k_2\cdot \mathbb{W} } p_1\,.
\end{equation}
This implies that we can write
\begin{eqnarray}
  |\varepsilon_3\rangle   &=&   e^{-k_2\cdot \mathbb{W} } |\varepsilon'_1\rangle \,,\qquad \,\,\qquad n=3 \\
  |\varepsilon_4\rangle   &=&   e^{-\mu M^2 (k_2+k_3)\cdot \mathbb{W} } |\varepsilon'_1\rangle \,,\quad n=4\label{eq:boost_4}
\end{eqnarray}
where $|\varepsilon'_1\rangle$ is a polarization state associated to $p_1=M u$. Thus we have the following QFT amplitudes
\begin{eqnarray}\label{eq:qftamp3}
    A^{S}_{3}= A^{0}_3 \times \langle \varepsilon'_1| e^{k_2\cdot \mathbb{W} } e^{-2 k_2 \cdot \mathbb{W}}  | \varepsilon_1 \rangle = A^{0}_3 \times \langle \varepsilon'_1| e^{-k_2\cdot \mathbb{W} }  | \varepsilon_1 \rangle \,,
    \end{eqnarray}
and
\begin{eqnarray}\label{eq:qftamp4}
    A^{S}_{4}=A^{0}_4 \times \langle \varepsilon'_1 | e^{\mu M^2 (k_2+k_3)\cdot \mathbb{W} } e^{2(w - k_2)\cdot \mathbb{W}} | \varepsilon_1 \rangle \,.
\end{eqnarray}
The constraint $u\cdot \mathbb{W}=0$ implies that the Pauli-Lubanski operator $\mathbb{W}^\mu$ only yields three independent operators. In the rest frame of $u^\mu$ they satisfy $[\mathbb{W}^i,\mathbb{W}^j]=i\epsilon^{ijk}\mathbb{W}_k$, or covariantly

\begin{equation}\label{eq:a-commutator}
    [\mathbb{W}^\mu,\mathbb{W}^\nu]=M^{-1} \mathbb{S}^{\mu \nu}=i\epsilon^{\mu \nu \rho \sigma}\mathbb{W}_\rho u_\sigma \,.
\end{equation}
In eq. \eqref{eq:qftamp3} only the combination $k_2 {\cdot}\mathbb{W}$ appears. Furthermore, note that in this case the boost component $e^{k_2 \cdot \mathbb{W}}$ commutes with the amplitude $e^{-2 k_2\cdot \mathbb{W}}$. This is not the case for eq. \eqref{eq:qftamp4} where indeed all three combinations $k_2{\cdot }\mathbb{W},k_3{\cdot } \mathbb{W}, w{\cdot } \mathbb{W}$ appear and do not commute among each other. As the spin is the only quantum number available, we assume that in general, these combinations span a basis of operators in the space of states associated with $u^\mu$, namely $|\varepsilon_1\rangle, \langle \varepsilon'_1|$.

\subsection{Classical Limit and Crossing Symmetry}\label{sec:classical_limit}

As argued in the previous section, the operator $\mathbb{O}$ in the contraction $\langle \varepsilon'_1 |\mathbb{O}|\varepsilon_1\rangle$ can be attributed a classical nature, that is $O\equiv\langle\mathbb{O}\rangle$. This requires the classical limit briefly mentioned in  \cref{sec:kinematics} and extensively studied in \cite{Bautista:2021wfy}. We note that the three-point amplitude \eqref{eq:qftamp3} is invariant under such limit

\begin{equation}
\label{eq:classical3pt}
    \langle A_3^S\rangle  = \langle A_3^0\rangle e^{-k_2\cdot a},
\end{equation}
where we have used $a^\mu=\langle\mathbb{W}^\mu\rangle$. For the  four-point case, from \eqref{eq:qftamp4} we obtain 
\begin{equation}
    w^\mu,k_2^\mu,k_3^\mu\sim \hbar \,, \quad \mathbb{W}^\mu \sim 1/\hbar \,.
\end{equation}
where the scaling of $w^\mu$ follows from its definition \eqref{eq:wdef}. Together \eqref{eq:a-commutator} this implies

\begin{equation}
    [(k_2+k_3)\cdot \mathbb{W} , (w - k_2)\cdot \mathbb{W}] \sim \hbar 
\end{equation}
i.e. the exponents of \eqref{eq:qftamp4} commute in the classical limit. Furthermore, from the explicit expression in \cite{Guevara:2019fsj} we see that $\mu M^2 = 1+\mathcal{O}(\hbar)$, hence the limit of \eqref{eq:qftamp4} becomes

\begin{eqnarray}\label{eq:qftamp4clas}
 A^{S}_{4} =A^{0}_4 \times    \langle \varepsilon'_1|e^{(2w +k_3 - k_2)\cdot \mathbb{W}}|\varepsilon_1\rangle +\mathcal{O}(\hbar)  \Longrightarrow \langle A^{S}_{4} \rangle =\langle A^{0}_4  \rangle \times  e^{(2w +k_3 - k_2)\cdot a}  \,.
\end{eqnarray}
The result for the  classical amplitude agrees with the one obtained in \cite{Aoude:2020onz} from Heavy Particle EFT. This is expected since as we have argued in \cite{Bautista:2021wfy},  the limits $\hbar \to 0$ and $M\to \infty$ are equivalent.

Note that in the last step of \eqref{eq:qftamp4clas} we have stripped off the polarization states $|\varepsilon_1\rangle, |\varepsilon'_1\rangle$. As a consistency check, one may ask if the same result is obtained if we start from the antichiral amplitude $ A_n^{\textrm{antichir.}}$ as defined in \eqref{eq:chiral-to-antichiral}. This was observed in \cite{Guevara:2019fsj} for $n=3$ but now we show it holds in general. To see this we need the following identity, which we derive in  \cref{ap:chiral}:

\begin{equation}\label{eq:chiralityflip}
    A_n^{\textrm{antichir.}} = e^{2K\cdot \tilde{\mathbb{W}}} A_n^{\textrm{chir.}|_{\mathbb{W}\to\tilde{\mathbb{W}}}} +\mathcal{O}(\hbar)
\end{equation}
Here  $\tilde{\mathbb{W}}^\mu$ corresponds to the Pauli-Lubanski operator acting on antichiral states $|\varepsilon_i]$:

\begin{equation}\label{spinvecac}
    \tilde{\mathbb{W}}^{\mu} :=\frac{1}{2m} \epsilon^{\mu \nu \rho \sigma } u_{\nu} \tilde{\mathbb{J}}_{\rho \sigma}
\end{equation}
Consequently, the superscript $\mathbb{W}\to\tilde{\mathbb{W}}$ means that the operator $\mathbb{W}^\mu$ in $A_n^{\textrm{chir.}}$ must be formally replaced by $\tilde{\mathbb{W}}^\mu$. We shall omit the superscript in the classical limit and assume that both $a^\mu = \langle\mathbb{W}^\mu\rangle$ and $\tilde{a}^\mu = \langle\tilde{\mathbb{W}}^\mu\rangle$ yield the same  interpretation as classical spin vector. The momentum $K$ is the total momentum transfer, i.e. $K=k_2$ for $n=3$ and $K=k_2+k_3$ for $n=4$. Now, applying the classical prescription on the chiral basis gives
\begin{eqnarray}\label{eq:chirclaslim}
    A_n^S&=&\langle \varepsilon_n |  A_n^{\textrm{chir.}} |\varepsilon_1 \rangle \nonumber \\
    &=& \langle \varepsilon'_1 |  e^{K{\cdot} \mathbb{W}} A_n^{\textrm{chir.}} |\varepsilon_1 \rangle   +\mathcal{O}(\hbar)  \longrightarrow \langle A_n^S\rangle = e^{K{\cdot} a} \langle A_n^{\textrm{chir.}} \rangle \,,
\end{eqnarray}
whereas on the antichiral basis, we obtain 
\begin{eqnarray}
    A_n^S&=&[\varepsilon_n |  A_n^{\textrm{antichir.}} |\varepsilon_1] \nonumber \\
    &=& [\varepsilon_n |  e^{2K\cdot \tilde{\mathbb{W}}} A_n^{\textrm{chir.}|_{\mathbb{W}\to \tilde{\mathbb{W}}}} |\varepsilon_1]  +\mathcal{O}(\hbar) \nonumber \\
        &=& [\varepsilon'_1 | e^{-K\cdot \tilde{\mathbb{W}}} e^{2K\cdot \tilde{\mathbb{W}}} A_n^{\textrm{chir.}|_{\mathbb{W}\to\tilde{\mathbb{W}}}} |\varepsilon_1]  +\mathcal{O}(\hbar) \,.
\end{eqnarray}
which after stripping off the antichiral states becomes
\begin{equation}\label{eq:ansclas}
    \langle A_n^S \rangle = e^{K\cdot a} \langle A_n^{\textrm{chir.}}\rangle \,.
\end{equation}
Crucially, the same result as \eqref{eq:chirclaslim} is obtained due to the fact that the exponent $K\cdot \mathbb{W}$ has different sign in the chiral and antichiral boosts. This reflects that $e^{\pm K{\cdot} \mathbb{W}}$ is indeed a boost and not a little group transformation.

We can use \eqref{eq:chiralityflip} to derive a new classical constraint for $n=4$ that follows essentially from Crossing symmetry of the two gravitons. For this, we first introduce some notation. Let us recall definition \eqref{eq:chiral-to-antichiral} and further introduce the polynomial function $P_\xi$, via

\begin{equation}\label{eq:fform}
    A_4^{\textrm{chir.}}= A_4^0\times P_\xi (k_2\cdot a,k_3 \cdot a, w\cdot a) \,.
\end{equation}
i.e. for any spin the amplitude operator can be expanded in terms of $k_2\cdot a,k_3\cdot a,w\cdot a$, as observed in the previous subsection. As all helicity dependence of the gravitons is encoded in $A_4^0$,  the function $P_\xi$ can only further depend on kinematic invariants. The only combination with a non-vanishing classical limit is the optical parameter $\xi$ of \eqref{eq:xidef}, which has the 
  advantage of exhibiting Crossing symmetry. In Appendix \ref{ap:bose} we show that the classical amplitude should satisfy the  following constraint 
\begin{equation}\label{eq:a4sbox}
   \boxed{ \langle A_4^
   S\rangle = \langle A_4^0\rangle e^{(k_2+k_3)\cdot a}  P_\xi (k_2\cdot a,k_3 \cdot a, w\cdot a) = \langle A_4^0\rangle e^{-(k_2+k_3)\cdot a}  P_\xi (-k_3\cdot a,-k_2 \cdot a, w\cdot a) }\,,
\end{equation}
which is the statement that the classical amplitude $\langle A_4^S \rangle$ is symmetric under the exchange of $k_2\cdot a \leftrightarrow -k_3\cdot a$. Note that this is trivially fulfilled by the exponentiated form \eqref{eq:qftamp4clas}. In general, this will place a strong constraint on the general spin amplitude.

In the remaining part of the section, our aim will be to provide the function $P_\xi$ for arbitrary spins. We constrain only its classical limit, i.e. we can always add (Crossing symmetric) combinations of kinematic invariants that vanish as $\hbar \to 0$. However, assuming $P_\xi$ only depends on $\xi$ provides a trivial quantum completion of the scattering amplitude.

\subsection{General Compton Amplitudes}\label{Sec:General Compton Amplitudes}
Our objective here  is  to provide a generic form of $A_4^S$ for arbitrary spins. Our approach will be based on imposing three-point factorization as given by the minimal coupling amplitudes \eqref{3ptex}. Working in the strict classical limit we seek an ansatz of the form

\begin{equation}\label{eq:ansatzspin}
    \langle A_4^S \rangle = \langle A_4^0\rangle \times \left(e^{(2w+k_3-k_2)\cdot a} + P_\xi (k_2\cdot a,-k_3 \cdot a, w\cdot a) \right )_{2S}\,.
\end{equation}
On the right-hand side, we have written the subscript $2S$ to simply emphasize that both functions will be truncated at order $a^{2S}$, being effectively polynomials for finite spin quantum number. Furthermore, it follows from \eqref{eq:a4sbox} that the polynomial $P_\xi$ must be symmetric in its first two entries $P_\xi(\alpha,\beta,\gamma)=P_\xi(\beta,\alpha,\gamma)$. 

Locality and unitarity constraints are implemented as follows: As it turns out, the exponential form \eqref{eq:4ptexpdef}, which yields the first term in \eqref{eq:ansatzspin}, contains the right three-point factorization for  arbitrary spin. This is easy to see following the same arguments made below \eqref{eq:4ptexpdef}, which persist even after the classical limit. For instance, in the s- or u- channels we have $e^{(2w+k_3-k_2)\cdot a} \to e^{k_3\cdot a} e^{-k_2\cdot a}$, etc...

Because of the above, we impose $\langle A_4 ^0 \rangle \times P_\xi$ not to have a pole in the physical factorization channels. Additionally, starting at order $a^5$ we require $\langle A_4 ^0 \rangle \times P_\xi$ to cancel the unphysical pole  $1/(u\cdot \epsilon_2)$ which appears from the exponential term due to the $w$ vector.\footnote{It follows from eqs. \eqref{eq:1pxi},\eqref{eq:xidef} that such singularity corresponds to backward scattering $\xi^{-1}\to -1$ or $\theta\to \pi$, as argued in \cref{sec:classical_cross_section}. The fact that the amplitude is expected to be finite in this limit has been emphasized in \cite{Dolan:2008kf,PhysRevD.16.237} from classical considerations.} More precisely, our strategy is to Laurent-expand in $\xi$:

\begin{equation}
    P_\xi (k_2\cdot a, -k_3 \cdot a, w\cdot a) =
    \sum_m \xi^m p^{(m)} (k_2\cdot a, -k_3 \cdot a, w\cdot a) \,.
\end{equation}
This can be thought of as a perturbative expansion away from the Eikonal $\xi\to\infty$. Now the polynomials $p^{(m)}$ only depend on spin operators. We then implement the following considerations:

\begin{enumerate}
    \item From definitions \eqref{eq:wdef} and \eqref{eq:xidef}, a pole in $\xi$ must be canceled via the spin operator $(w\cdot a)^2$ (away from the strict classical limit, the Crossing symmetry constraint requires us to replace $(w\cdot a)^2\to -w\cdot a\, w'\cdot a$, see definition \eqref{eq:a4antic}). Note also that $\langle A_4^0 \rangle$ contains a simple pole in $\xi$, as given explicitly in \eqref{eq:a40xi}.     This yields
    \begin{equation}
        p^{(m)} \propto (w\cdot a)^{2-2m} \,\quad \textrm{for}\quad m\leq 1 \,.
    \end{equation}
    \item A pole in $t=\langle 23 \rangle [23]$ yields two different factorization channels $\langle 23\rangle \to 0$ and $[23]\to 0$. To cancel such pole we 
 will again employ the observation \eqref{eq:obst1} for each of these branches. Hence, each power of $1/t$ must be cancelled by the combination $(w\cdot a- k_2\cdot a)(w\cdot a + k_3 \cdot a)$, which trivially fulfills the crossing constraint \eqref{eq:a4sbox}. From \eqref{eq:xidef} we note that each power of $\xi$ contains a pole in $t$. Moreover, recall that $\langle A_4^0 \rangle$ also contains one such pole. Hence,
    
    \begin{equation}
        p^{(m)} \propto (w\cdot a- k_2\cdot a)^{m+1}(w\cdot a + k_3 \cdot a)^{m+1}\,\quad \textrm{for}\quad m\geq -1\,,
    \end{equation}
    
    \item The unphysical pole $u\cdot \epsilon_2\propto \langle 2 |u|3]$ is contained in $w\cdot a$. To cancel this pole we invoke the following useful identity
    \begin{equation}\label{eq:1pxi}
       \frac{\langle 2|u|3]\langle 3|u|2]}{t}= 1+\xi  \,.
    \end{equation}
    as well as (using the definition in \eqref{eq:tildew})
    
\begin{equation}\label{eq:wpluswtilde}
    w\cdot a + \tilde{w}\cdot a = \frac{\xi}{1+\xi} (k_2\cdot a - k_3\cdot a)\,,.
\end{equation}

The latter relation reflects that the conjugate operator $\tilde{w}\cdot a$ indeed can be expressed in terms of the basis $\{w\cdot a, k_2\cdot a,k_3\cdot a\}$. Since $\tilde{w}$ contains $\langle 3|u|2]$ in the denominator, in principle both terms on the LHS contribute to the pole as $\xi \to -1$. However, we will take $\langle 3|u|2]$ and $\xi$ to be independent variables (we can solve for $\langle 2|u|3]$ from \eqref{eq:1pxi}). The relation \eqref{eq:wpluswtilde} then reveals that as $\xi \to -1$ the term $\tilde{w}\cdot a$ drops and our basis $\{w\cdot a, k_2\cdot a,k_3\cdot a\}$ becomes degenerate!

The strategy is then to use \eqref{eq:wpluswtilde} to solve for $w\cdot a$ in terms of a new (non-degenerate) basis $\{\tilde{w}\cdot a, k_2\cdot a,k_3\cdot a\} $. Only then we demand the cancellation of the pole in $(1+\xi)$. Recall that $\langle A_4^0\rangle$, eq. \eqref{eq:a40xi}, has a fourth order zero in $\langle 2 |u|3]\propto 1+ \xi$.  Note further that since we start with an ansatz that does not contain $\tilde{w}\cdot a$, we are guaranteed that the pole in $\langle 3|u|2]$ is spurious, even though it may appear explicitly in some of the terms.

\item There is a final caveat to the above construction, which stems from the following identity in the classical limit 

\begin{equation}\label{eq:a2identity}
     -\frac{(s-M^2)(u-M^2)}{4M^2} a^2 \approx \omega^2 a^2 \approx  \xi(w\cdot a - k_2 \cdot a) (w\cdot a +k_3 \cdot a) +(w\cdot a)^2
\end{equation}
   (recall $\omega$ was defined in \eqref{eq:sandtclassical}). As expected, this shows that the operator $-a^2$ can indeed be expanded in the basis. However, one can check that when acting on the states \eqref{basi1} $-a^2$ is positive-definite: In fact, it corresponds to the quadratic Casimir of $su(2)$, as argued in Appendix \ref{sec:spinningspheroidalharmonics}. Thus, it is natural to introduce the operator $|a|=\sqrt{-a^2}$ by defining its action on spin-s states. Now, because of the quadratic nature of the relation \eqref{eq:a2identity}, we find that $\omega 
    |a|$ is indeed \textit{linearly} independent from $\{w\cdot a,k_2\cdot a,k_3\cdot a\}$. This means it can be included in our polynomial expansion, but only at the linear order.\footnote{In principle one can introduce (a crossing-symmetric version of) the combination $(s-M^2)|a| (w\cdot a)$ to cancel a pole in $\xi$, which would modify Constraint 1. For instance, at $\mathcal{O}(a^{4,5})$, one could add the extra contact terms $ct_4\sim \langle A_4^0\rangle \omega |a|w{\cdot}a(w{\cdot}a-k_2{\cdot}a)(w{\cdot}a+k_3{\cdot}a)$ and $ct_5\sim ct_4(k_2{\cdot}a-k_3{\cdot}a) $ respectively.
    However, we find that this is not required for the classical matching and does not change the conclusions below, so we will ignore them and stick to the prescription given by Constraint 1 above.  } 

     \end{enumerate}  
    
    Inclusion of the $|a|$ operator is what allows us to  match  the full classical computation including \textit{both} conservative and dissipative contributions. However, in \ref{sec:mathcing_comtpon} we will present an alternative in which the spin norm $|a|$ is treated as a c-number rather than an operator. Nicely, allowing the coefficients of the ansatz to be non-polynomial functions of $|a|$ yields also an exact matching for each monomial in $\{w\cdot a,k_2\cdot a,k_3\cdot a\}$. Furthermore, we will argue that while conservative contributions are captured entirely by operators in  the $\{w\cdot a,k_2\cdot a,k_3\cdot a\}$ basis, dynamical effects at the BH horizon (see e.g.\ \cite{Goldberger:2020wbx,Goldberger:2019sya}) can be accounted for by allowing $|a|$ terms, as we will see in \cref{sec:mathcing_comtpon}.     
    
Continuing with our discussion, constraints 1 and 2 can be imposed right away, leading to the expansion 

\begin{align}\label{eq:ansatz}
    P_\xi =& \sum_{m=0}^2 \xi^{m-1} (w\cdot a)^{4-2m}(w\cdot a- k_2\cdot a)^m (w\cdot a + k_3 \cdot a)^m r^{(m)}_{|a|}(k_2\cdot a, -k_3 \cdot a, w\cdot a) \nonumber \\ 
   & +\sum_{m=0}^{\infty} \left[ \frac{(w\cdot a)^{2m+6}}{\xi^{m+2}} p^{(m)}_{|a|}(k_2\cdot a, -k_3 \cdot a, w\cdot a) \right. \nonumber \\
    &\left. + \xi^{m+2} (w\cdot a- k_2\cdot a)^{m+3}(w\cdot a + k_3 \cdot a)^{m+3}\,q^{(m)}_{|a|}(k_2\cdot a, -k_3 \cdot a, w\cdot a) \right]
\end{align}
where $p^{(m)}_{|a|},q^{(m)}_{|a|},r^{(m)}_{|a|}$ are multivariable polynomials symmetric in their first two arguments, which also include a linear correction in $\omega |a|$. Note that the infinite sum is $\mathcal{O}(a^6)$.  Let us for the moment focus here on the first line. Crucially, this is $\mathcal{O}(a^4)$: \textit{This means that there are no effective operators that survive the classical limit up to order $a^3$, or equivalently, for particles of spin $s<2$}. If we assume that the Kerr background can be effectively matched to a certain Compton amplitude, we can already conclude that the minimal coupling amplitudes \eqref{ampli} will indeed match up to order $a^3$. We have seen this is the case for polar scattering at linear order in the previous section and will confirm it in the general (non-polar) case in the next section up to $a^6$ order. 

After using constraints 1-4, we can easily parametrize the polynomials $r^{(m)}_{|a|}$, $p^{(m)}_{|a|}$, and $q^{(m)}_{|a|}$, leading to the complete result for $P_\xi$ up to order $a^6$:\footnote{In a crossing-symmetric fashion, $\omega \approx \sqrt{(s-M^2)(M^2-u)/4M^2}\approx(s-M^2)/2M$. 
Extensions to higher orders in spin follow analogously.}
\begin{equation}
   \begin{split}r_{|a|}^{(m)} & =c_{1}^{(m)}+c_{2}^{(m)}(k_{2}{\cdot}a-k_{3}{\cdot}a)+c_{3}^{(m)}w{\cdot}a+c_{4}^{(m)}|a|\omega\\
 & \quad+c_{5}^{(m)}(w{\cdot}a-k_{2}{\cdot}a)(w{\cdot}a+k_{3}{\cdot}a)\\
 & \quad+c_{6}^{(m)}(2w{\cdot}a-k_{2}{\cdot}a+k_{3}{\cdot}a)w{\cdot}a\\
 & \quad+c_{7}^{(m)}(2w{\cdot}a-k_{2}{\cdot}a+k_{3}{\cdot}a)^{2}+c_{8}^{(m)}(w{\cdot}a)^{2}\\
 & \quad+c_{9}^{(m)}(k_{2}{\cdot}a-k_{3}{\cdot}a)|a|\omega+c_{10}^{(m)}w{\cdot}a|a|\omega+\mathcal{O}(a^{3})
\end{split}
\end{equation}
\begin{equation}
    p_{|a|}^{(m)}=d_{1}^{(m)}+\mathcal{O}(a)\,,\qquad q_{|a|}^{(m)}=f_{1}^{(m)}+\mathcal{O}(a)\,.
\end{equation}
Constraint 3 imposes  the relation shown in the second column of \cref{tab:Teukolskysolutions}, at the indicated  order in spin.
The remaining free coefficients will be fixed from solutions to the Teukolsky equation.

\subsection{Matching to Teukolsky computation}\label{sec:matchTA}
After imposing constraints 1.-4.\ above, the remaining free coefficients  of the Compton ansatz can  be fixed  by matching to the full non-polar GW scattering off a Kerr background.
However, before we can implement such matching   we need to rewrite our ansatz in a suitable language: the spinning partial-wave basis, as we did  for the lower helicity cases in \cite{Bautista:2021wfy}. This is a technical task whose details we postpone to be discussed in \cref{Sec:PWScattering} together with Appendix \cref{sec:spinningspheroidalharmonics}, using massive spinor-helicity. 
For the moment in \cref{tab:Teukolskysolutions} we just   summarize our findings from the   matching procedure.

\begin{table}
\begin{centering}
\begin{tabular}{|c|c|c|c|}
\hline 
Spin & Spurious-pole & Free Coeffs.&Teukolsky Solutions\tabularnewline
\hline 
\hline 
$a^{4}$ & & $c_{1}^{(i)},\,\,i=0,1,2$ & $c_{1}^{(i)}=0,\,\,i=0,1,2$\tabularnewline
\hline 
\hline 
$a^{5}$ &$c_{3}^{(2)}=4/15-c_{3}^{(0)}+c_{3}^{(1)}$ &$ \begin{aligned}
    c_{2}^{(i)}&,\,\,i=0,1,2 \\
    c_{3}^{(i)}&,\,\,i=0,1\\
    c_{4}^{(i)}&,\,\,i=0,1,2       
\end{aligned}$
& $\begin{aligned}c_{2}^{(i)} & =0,\,\,i=0,1,2\\
c_{3}^{(0)} & =\alpha\frac{64}{15}\,,\,c_{3}^{(1)}=\alpha\frac{16}{3}\,,\\
c_{3}^{(2)}&=\frac{4}{15}(1+4\alpha)\,,\\
c_{4}^{(0)} & =\eta\alpha\frac{64}{15}\,,\\
c_{4}^{(1)}&=\eta\alpha\frac{16}{5}\,,\,c_{4}^{(2)}=\eta\frac{4}{15}
\end{aligned}
$\tabularnewline
\hline 
\hline 
$a^{6}$ & $\begin{aligned}
c_{10}^{(2)} & =c_{10}^{(1)}-c_{10}^{(0)}\\
d_{1}^{(0)}=&-\frac{8}{45}\\
&+\sum_{j=5}^7\sum_{i=0}^{2}(-1)^i c_j^{(i)}\\
f_{1}^{(0)}  =&\frac{4}{45}+c_6^{(0)}-c_6^{(1)}\\
&+\sum_{i=0}^2(-1)^ic_8^{(i)}
\end{aligned}
$ & $
\begin{aligned}
     c_{5}^{(i)}&,\,\,i=0,1,2 \\
    c_{6}^{(i)}&,\,\,i=0,1,3\\
    c_{7}^{(i)}&,\,\,i=0,1,2 \\
    c_{8}^{(i)}&,\,\,i=0,1,3\\
    c_{9}^{(i)}&,\,\,i=0,1,2  \\
     c_{10}^{i}&,\,\,i=0,1
\end{aligned}
$

& $\begin{aligned}c_{j}^{(i)} & =0,\,\,i=0,1,2,\,\,j= 5,7\\
c_{6}^{(0)} & =\alpha\frac{128}{45}\,,\,c_{6}^{(1)}=\alpha\frac{32}{9}\,,\\
c_{6}^{(2)}&=\frac{8}{45}(1{+}4\alpha)\,,
c_{8}^{(0)}  ={-}\alpha\frac{512}{45},\,\\
c_{8}^{(1)}&={-}\alpha\frac{160}{9},\,c_{8}^{(2)}={-}\frac{16}{45}(1{+}19\alpha),\\
c_{9}^{(0)} & =-\eta\alpha\frac{128}{45}\,,\,c_{9}^{(1)}=-\eta\alpha\frac{32}{15}\,,\\
c_{9}^{(2)}&=-\eta\frac{8}{45},\\
c_{10}^{(0)} &=-\eta\alpha\frac{256}{45},\\
c_{10}^{(1)}&=-\eta\alpha\frac{352}{45},c_{10}^{(2)}=-\eta\alpha\frac{32}{15}\\
d_{1}^{(0)} & =0\,,\,f_{1}^{(0)}=-\frac{4}{45}(1+4\alpha)
\end{aligned}
$\tabularnewline
\hline 
\end{tabular}
\caption{Second column:. Spurious pole cancellation constraints on the Compton ansatz  at the given order in spin. The third column indicates the number of free contact terms after imposing the constraint of the second column. 
Operators $|a|$ were not considered previously  in  \cite{Aoude:2022trd}. Up to spin $a^5$, the remaining number of free coefficients coincide with the counting in this  reference. At $a^6$ the authors mention there are $19$ free parameters, out of which, only 12 are linearly independent.
The last column shows the values for the free coefficients that  match the  full Teukolsky solutions. Here $\alpha=1$, is a coefficient tracking  non-rational (digamma functions) contributions in the Teukolsky solutions, and $\eta = 0,\pm1$, is a parameter that keeps track of the conservative ($\eta=0$), and absorptive ($\eta=\pm1$) pieces of the  scattering amplitude (see \cref{Sec:PWScattering} for a detail explanation of the matching procedure). }
\label{tab:Teukolskysolutions}
\par\end{centering}
\end{table}

We notice up to $a^4$,   no contact terms are allowed by the Kerr BH, and the minimal coupling exponential from \eqref{eq:qftamp4clas} is enough to  capture the spin dynamics. At orders, $a^{5,6}$, we encounter three sets of solutions given by   $\eta=0,\pm1$. The first set ($\eta=0$) corresponds to the extraction of the \textit{conservative} piece of the   amplitude, where BH horizon absorption is removed before analytically continuing the BHPT results  from $a^\star=\frac{a}{GM}\le1$ to $a^\star\gg1$ (details are provided in \cref{Sec:PWScattering}). In this case, the result is independent of the prescription taken when analytically extending Teukolsky solutions  through  the singular point    $a^\star=1$  in the complex $a^\star$ plane. 
The other two sets ($\eta=\pm1$) keep track of dissipative contributions in  the scattering problem,   where the   sign is dictated by the prescription taken for the analytic continuation; the positive sign corresponds  to extending 
 the BHPT solutions by going above the singular point in the complex $a^\star$ domain, whereas the negative sign corresponds to extensions by going below the singular point. We will extend on this in \cref{Sec:PWScattering}. As a final remark of this subsection,  notice that in the conservative sector dropping the non-rational contributions ($\alpha=0$) sets  to zero all the contact terms of the Compton ansatz, except those strictly  needed to cancel the unphysical pole of the BCFW exponential. For the  case of scattering of scalar waves off Kerr \cite{Bautista:2021wfy},  removal of the non-rational terms  provided Teukolsky solutions that matched precisely the Born amplitudes (see eq. (3.8) in \cite{Bautista:2021wfy}), independent of contact term contributions. The  amplitude with no contact terms after the  unphysical pole is removed is then the gravitational analog of the Born amplitudes of the lower helicity cases. Let us however remark even for the conservative case, $\alpha$-contributions   should be kept in order for the amplitude to correctly describe the interaction of the wave with the Kerr BH (this includes the contact term contributions matching the  digamma functions for the scalar case discussed in Part I).

\subsection{Polar Scattering Revisited}
In light of the above reformulation, we can  revisit the  polar scattering scenario  introduced in Section \ref{sec:kinematics}. Because the spin is aligned with incoming momenta the spin basis $\{w\cdot a,k_2 \cdot a, k_3\cdot a,|a|\}$ becomes essentially one-dimensional. In fact, replacing $a^{\mu}$ by its classical value in polar scattering, $a^{\mu}=(0,0,0,a_z)$,
we  can easily show (see also \eqref{eq:expdecp})
\begin{equation}
k_2\cdot a = w\cdot a = - \frac{\xi}{\xi+2 }k_3\cdot a = \omega a_z \,, \,\,\,|a|=a_z\,.    
\end{equation}
Then the general ansatz \eqref{eq:ansatzspin} takes the following form
\begin{equation}
     \langle A_4^s \rangle_{\textrm{polar}} = \langle A_4^0\rangle \times \left[e^{-2\omega a_z/\xi} + P_\xi \left(\omega a_z,\frac{(\xi+2)\omega a_z}{\xi} , \omega a_z\right) \right ]
\end{equation}
The crucial observation is that the exponent in the first term is now regular as $\xi\to 1$, which previously corresponded to the backward ($\theta=\pi$) unphysical singularity; Hence, the exponential form of the Compton amplitude gives a sensible result for polar scattering as already discussed in \cref{sec:classical_cross_section}.
Using the solution above up to order  $a^6$, we can explicitly check that  for the Kerr Black Hole
\begin{equation}
    P_{\xi}\left(\omega a_{z},\frac{(\xi+2)\omega a_{z}}{\xi},\omega a_{z}\right)=-\frac{64}{15}\alpha(\eta-1)\sin^{2}(\theta/2)a_{z}^{5}\omega^{5}\left[1+\frac{2}{3}a_{z}\omega(3+\cos\theta)\right]+\mathcal{O}(a^{7})
 \,.
\end{equation}
This means that -- up to $a^6$ -- dropping the digamma contact contributions (setting $\alpha\to0$) makes the  polar scattering amplitude coincide with the BCFW exponential $e^{-2\omega a_z \sin^2(\theta/2)} $. As already mentioned, this exponential is the gravitational analog of the Born amplitudes for the $h<2$ cases. On the other hand, if the digamma contributions are kept, in the \textit{conservative} case  ($\eta=0$), the exponential receives a  modification caused by the digamma terms. (For the $h=0$ case, no contact modifications survive under the polar limit for the \textit{conservative} amplitude.) This extra contribution could be removed by including dissipative terms with  $\eta=1$.

In \cref{sec:apc2} we elaborate on polar scattering. In particular, using massive spinor-helicity variables, it is shown that the expansion in \textit{spinning spherical harmonics} truncates at each order in spin. This expansion is the topic of the next section.

\section{Classical wave scattering in Kerr spacetime}\label{Sec:PWScattering}

In this section we first recap the tools of Black Hole Perturbation Theory (BHPT) for the Kerr metric, and then proceed to compute a classical wave scattering amplitude to match the QFT ansatz as promised. This calculation of the BHPT amplitude is based on the one already presented in Part I \cite{Bautista:2021wfy} for a scalar wave, but involves several new ingredients associated to the helicities carried by Gravitational Waves (GW).

In Kerr spacetime, the differential cross-section for the scattering of a plane gravitational wave can be expressed as
\begin{align}
\frac{d\sigma}{d\Omega}=|f(\vartheta,\varphi)|^2+|g(\vartheta,\varphi)|^2,
\end{align}
where $f$ and $g$ are respectively the complex helicity-preserving and helicity-reversing scattering amplitudes. Using a partial wave decomposition,  they are given by the expressions
\begin{align}
     f(\vartheta,\varphi)&=\sum_{l=2}^{\infty} \sum_{m=-\infty}^{\infty}{}_{-2}S_{lm}(\gamma,0;a\omega){}_{-2}S_{lm}(\vartheta,\varphi;a\omega)f_{lm}, \label{Eq:fKerrGeneric}\\
     g(\vartheta,\varphi)&=\sum_{l=2}^{\infty} \sum_{m=-\infty}^{\infty}{}_{-2}S_{lm}(\gamma,0;a\omega){}_{-2}S_{lm}(\pi-\vartheta,\varphi;a\omega)g_{lm} \label{Eq:gKerrGeneric}
\end{align}
where ${}_{s}S_{lm}$ are the spin-weighted spheroidal harmonics, $\gamma$ is the angle between the incoming wave vector and the axis of rotation of the Kerr BH, which we consider non-zero in general  \cite{futterman88,Dolan:2008kf,Glampedakis:2001cx,Stratton:2020cps}. The first copy  of the  harmonics in these expressions  follows from the harmonic decomposition of the plane wave into the basis of spin-weighted spheroidal harmonics, whereas the second copy is the usual separation of variables ansatz for the Teukolsky scalar.
The amplitude modes are given by
\begin{align}
    f_{lm}&=\frac{2\pi}{i\omega}\sum_{P=\pm1}\left(e^{2i\delta^{P}_{lm}}-1\right), \label{eq:f_deff}\\
    g_{lm}&=\frac{2\pi}{i\omega}\sum_{P=\pm1}P(-1)^{l+m+2}\left(e^{2i\delta^{P}_{lm}}-1\right)\,,\label{eq:g_deff}
\end{align}
where $\delta^{P}_{lm}$ are the phase shifts, which for the gravitational case have the explicit form
\begin{equation}\label{eq:phase_shift}
    e^{2i\delta^{P}_{lm}}=(-1)^{l+1}\frac{\mathcal{C}_{lm}+12iM\omega P}{16\omega^4}\frac{B_{lm\omega}^{\text{ref}}}{B_{lm\omega}^{\text{inc}}}\,
\end{equation}
where $\mathcal{C}_{lm}$ is the Teukolsky-Starobinsky constant given explicitly in \eqref{eq:tsconstant}, and the $B_{lm\omega}$ coefficients are extracted from the asymptotic solutions to the radial Teukolsky equation \cite{Sasaki:2003xr}. 
Computing the phase shifts is standardly done by solving the radial Teukolsky equation, as discussed in Part I \cite{Bautista:2021wfy}, and e.g. \cite{Dolan:2008kf}, which we shall not discuss here. The main remaining point to be mentioned is that the imposition of the physical boundary condition that the modes for the scattering problem are purely transmitting into the horizon (and nothing coming out) fully determines the phase shifts given by \eqref{eq:phase_shift}. We refer the reader to Appendix \ref{PW-App} for the discussion of the subtleties involved in decomposing a plane wave metric perturbation onto a harmonic basis. 

For the purposes of matching the gravitational Compton  amplitude of previous sections, we calculate the partial wave amplitudes in a long wavelength limit $\epsilon = 2GM\omega\ll1$\footnote{The factor of 2 in the definition of the dimensionless expansion parameter $\epsilon$ is essentially historical and matches what is commonly used in the BHPT literature.}. As discussed in Part I, for our computation it is crucially important that $0\leq a^\star<1$ when constructing the long wavelength expansion, so that we can use the tools of black hole perturbation theory. When  $a^\star>1$ the BH ceases to have a horizon and standard methods for solving the Teukolsky equation are not clearly defined. In this regime,  we are effectively left with a naked singularity. We expect however an analytic continuation  to the $a^\star\gg1$ region to provide sensitive results that can match the effective point particle description of the BH given by the classical limit of the QFT scattering amplitude.

Before proceeding to discuss explicit BHPT results, let us do the following observation: the explicit matching of the BHPT  to the QFT results carries with it a set of technical steps as we observed for the scalar case in Part I \cite{Bautista:2021wfy}. The reason is that in general, doing the infinite sums of the previous partial waves is an almost impossible task. The strategy is then to expand the QFT and BHPT amplitudes into a suitable basis of partial waves that makes  simple the comparison between the two. For our purpose, we find it simpler to do such a comparison on the basis of spin-weighted spherical harmonics ${}_{-2}Y_{lm}(\theta  ,\phi')$.  Here $(\theta,\phi')$ is an intermediate basis between the QFT basis $(\theta, \phi)$ \eqref{eq:kinematics},  and the BHPT basis $(\vartheta,\varphi)$ \eqref{Eq:fKerrGeneric}. 
Then, additional work has to be done in order to align the two results into the $(\theta,\phi')$ basis. Although at this stage this might seem like an unnecessary step to take, we will see writing the  amplitudes in this basis provides some advantages over the spin-weighted spheroidal basis \eqref{Eq:fKerrGeneric}, especially when it comes to studying the analytic continuation  of the BHPT results to the $a^\star\gg1$ region. In order to avoid unnecessary details of this procedure, we refer the reader to  Section $4.3$ of \cite{Bautista:2021wfy} where we   extensively explained how to do this bases alignment and show the relationship among the different coordinate system, so that  in this work we limit ourselves to  provide the final results. 

In the basis of  spin-weighted spherical harmonics ${}_{-2}Y_{lm}(\theta  ,\phi')$, scattering  amplitudes can be written as 
\begin{align}\label{eq:fylmex}
    f(\theta,\phi')&=\sum_{l=2}^{\infty} \sum_{m=-\infty}^{\infty}{}_{-2}Y_{lm}(\theta,\phi')f'_{lm}(\gamma)\,.
\end{align}
For the BHPT amplitudes, the mode functions $f'^{(\text{BHPT})}_{lm}(\gamma)\equiv f'_{lm}(\gamma)$ no longer have completely factorized out the dependence on $\gamma$.  Rather, these    functions take the  form 
\begin{equation}\label{eq:modesfBHPT}
f'^{(\text{BHPT})}_{lm}(\gamma)=\sum_{m'}D^{l*}_{m'm}(\gamma)\; f_{lm'}\,,
\end{equation}
where $f_{lm'}$ are the functions entering in \eqref{Eq:fKerrGeneric}. Here $D^{l*}_{mm'}$ is the (complex conjugate) Wigner $D$-matrix with Euler angles $(0,\gamma,0)$,
\begin{equation}
D^{l*}_{mm'}(\gamma)=D^{l*}_{mm'}(0,\gamma,0)=(-1)^{m'}\sqrt{\frac{4\pi}{2l+1}}{}_{-m'}Y_{lm}(\gamma,0)\,.
\end{equation} 

Analogously, for the QFT amplitude the mode functions $f'^{\text{QFT}}_{lm}(\gamma)\equiv f'_{lm}(\gamma)$ can be computed in two ways starting from the Compton ansatz \eqref{eq:ansatzspin}. A straightforward way is by employing integrals over the 2-sphere,
\begin{equation}\label{eq:compton_modes}
f'^{\text{QFT}}_{lm}(\gamma)=\int d\Omega'\,{}_{-2}Y^*_{lm}(\theta,\phi') \langle A_4(\gamma,\theta,\phi')\rangle,
\end{equation}
where in this basis,  the components of the  spin vector entering the amplitude need to be taken as  
\begin{equation}\label{eq:axayaz}
a_z=a\cos\gamma, \quad
 a_x=-a\sin\gamma\cos\phi', \quad
 a_y=- a\sin\gamma\sin\phi'\,.
\end{equation}
This projection however does not provide much intuition on the relation between the partial amplitudes $f_{lm}'$ and the contact operators $\{k_2\cdot a,k_3\cdot a,w\cdot a\}$ in our ansatz. As both carry the full angular dependence of the amplitude a direct correspondence is expected. This correspondence is established in \cref{sec:spinningspheroidalharmonics}, by exploiting a novel construction of harmonics using the spinor-helicity variables of the previous section.

In any case, agreement of the QFT and BHPT results means then that the equality
\begin{equation}\label{eq:comparison}\boxed{
    f'^{\text{QFT}}_{lm}(\gamma) = f'^{\text{BHPT}}_{lm}(\gamma)\,,}
\end{equation}
is satisfied for all values of $l,m$. In what follows we will focus our attention to analyze these mode functions, both in the BHPT and QFT approaches. 

\subsection{Analytic extension and anomalous behaviour}\label{sec:anomalous}
Following the notation of Part I, the BHPT amplitude modes can be decomposed into a piece containing the Newtonian term, an overall phase, and functions coefficients containing the dependence on the  BH spin, the PM parameter $\epsilon$, and the inclination angle $\gamma$
\begin{equation}
    f'^{\text{BHPT}}_{lm}(\gamma)= e^{i\Phi}\frac{\Gamma(l+1-i\epsilon)}{\Gamma(l+1+i\epsilon)}\beta_{lm}(a^\star, \epsilon,\gamma)\,,\label{Eq:phasetobeta}
\end{equation}
The key observation is that the mode functions have a low-energy decomposition\footnote{ Note that there is no  first-order term in $\epsilon$ as it cancels exactly, as first observed in \cite{Dolan:2008kf}.}
\begin{equation}\label{eq:beta_modes}
    \beta_{lm}(a^\star, \epsilon,\gamma)= 1+\beta_{lm}^{(2)}(\gamma)\epsilon^2+\beta_{lm}^{(3)}(\gamma)\epsilon^3+\beta_{lm}^{(4)}(\gamma)\epsilon^4+\cdots
\end{equation}
At each order $i$, in $\epsilon$, the mode coefficients are then exact functions of $a^\star$ (and $\gamma$). We explicitly obtained results up to $i=7$. Naively it might seem as if higher $\epsilon$ terms are also higher PM contributions. However, we need to recall $a^\star=a/GM$ entering in the mode coefficients also carries negative powers of $G$, therefore making the combination $z^n=(\epsilon a^\star)^n =(2a\omega)^n$, of order $\mathcal{O}(G^0)$, indeed  contributing to the tree-level amplitude. This was observed at linear order in spin in \cite{Dolan:2008kf}, and we have extended to all spin orders in Part I, and up to the sixth order in spin in this paper. 

To compare with QFT results, we need values for the mode  coefficients for $a^\star\gg1$. We achieve this by analytically extending our BHPT results into this domain. This extension, however, needs to be carried out in a careful way; the reason is the following: in general, the mode coefficients contain both a conservative and  an absorptive piece, given by the real and imaginary contributions respectively\footnote{It should be clear the angle $\gamma$ is real and therefore trigonometric functions of $\gamma$ are always real. }. The $a^\star\gg1$ extension mixes the two contributions, therefore, giving us a result containing both of them. To avoid this we have two options: If on the one hand one is interested in extracting  the purely conservative contribution of the classical amplitude, one can remove the absorptive contributions before analytic continuation, as we will explain below.  On the other hand, if we want to keep these absorptive pieces and match them to the certain operators in the Compton ansatz, we need to  carefully keep track of them while doing the   $a^\star\gg1$ extension;  we will show an explicit example below.  

The explicit expressions of the $\beta^{(i)}_{lm}(\gamma)$ coefficients, for all $l,m$, and   for $i\leq 5$, are  polynomials   of  $a^\star$ and have a unique analytic extension. Importantly, $\left(a^\star\right)^4$ is the highest power of the spin that appears in these cases and the functions are purely real, therefore containing only  conservative contributions, which perfectly match up to the fourth order of the expansion of the exponential in \eqref{eq:ansatzspin}. This in turn  fixes to zero the contact deformations at $a^4$, as indicated in \cref{tab:Teukolskysolutions}. 
These features remain true  for $i=6,7$ when $l>3$, matching as well the exponential part of the Compton amplitude \eqref{eq:ansatzspin}, once the unphysical pole is removed. Therefore, we expect the remaining contact terms in \eqref{eq:ansatzspin} to contribute only to  the lower $l=2,3$ harmonics; indeed the higher harmonics coefficients come mostly from the harmonic expansion of the t-channel pole.

In analogy to the analysis  for the case of scalar waves presented in Part I, for the gravitational case we also  find anomalous behavior for the expansion coefficients  \eqref{eq:beta_modes}, for  certain low values of $l$, namely  for $l=2,3$, as expected,  in $\beta_{lm}^{(6,7)}$. This anomaly comes from the presence of non-rational functions of $a^\star$ in the mode coefficients, together with the presence of absorptive pieces. For instance, for $l,m=2,-1$, at $a^5$, for the helicity preserving amplitude, $f_{lm}$, the function coefficients have the explicit form 
\begin{align}\label{eq:modes}
    \beta_{2,-1}^{(6)}(\gamma)= & \frac{\sqrt{\pi/5}\sin^{3}\gamma}{42247941120}\Big[-43659(12017+17775\cos(2\gamma))a^{\star5}- 1408264704\cos(\gamma) \,i a^{\star4}\hat{\kappa}\nonumber\\
 & -704132352\Big((1-2\cos\gamma)\psi^{(0)}(-ia^{\star}/\hat{\kappa})+(1+2\cos\gamma)\psi^{(0)}(ia^{\star}/\hat{\kappa})\Big)a^{\star3}\nonumber\\
  & -5633058816\Big(\sin^{2}(\gamma/2)\psi^{(0)}(-2ia^{\star}/\hat{\kappa})+\cos^{2}(\gamma/2)\psi^{(0)}(2ia^{\star}/\hat{\kappa})\Big)a^{\star3}\Big]\,,
 % \\ \beta_{2,1}^{(5)}=  & -\frac{\sqrt{\pi/5}\sin\gamma}{633719116800}\Big[-1309770\big(789+1936\cos(2\gamma)-1605\cos(4\gamma)\big)a^{\star5}\nonumber\\
 % & -21123970560\cos\gamma\sin^{2}(\gamma)i\hat{\kappa}a^{\star4}-84495882240\cos(\gamma)\sin^{4}(\gamma/2)\psi^{(0)}(-ia^{\star}/\hat{\kappa})a^{\star3}\nonumber\\
 % & -42247941120\Big(\cos^{4}(\gamma/2)(1-2\cos\gamma)\psi^{(0)}(ia^{\star}/\hat{\kappa})+\sin^{4}(\gamma/2)\psi^{(0)}(-ia^{\star}/\hat{\kappa})\Big)a^{\star3}\nonumber\\
 % & +337983528960\Big(\cos^{6}(\gamma/2)\psi^{(0)}(2ia^{\star}/\hat{\kappa})+\sin^{6}(\gamma/2)\psi^{(0)}(-2ia^{\star}/\hat{\kappa})\Big)a^{\star3}\Big]
\end{align}
where $\hat{\kappa}=\sqrt{1-a^{\star2}}$ and $\psi ^{(0)}(z)=\Gamma'(z)/\Gamma(z)$ is the digamma function. 
Here we have discarded terms irrelevant for the $a^\star\gg1$ expansion. 
Notably, these functions are complex for sub-extremal ($a^\star<1$) Kerr BHs and have a singular point at extremality $a^\star=1$. As already mentioned,  the finding of  imaginary modes in the  amplitude signals that the interaction with the Kerr BH is in general not conservative. Following  \cite{Dolan:2008kf}, we interpret these imaginary  contributions as horizon absorption effects. Our prescription  to extract the conservative part of the amplitude is the same  we used in  Part I for the scattering of scalar waves, where absorption was removed by taking the real part of the phase shifts before extending the BHPT results to the $a^\star\gg1$ region. In practice, this is achieved by averaging the mode coefficients \eqref{eq:modes}, with their respective complex conjugate. After this is done, the $a^\star\gg1$ extension is performed. In this paper, however, we aim   more generally to match the full Teukolsky solutions to the Compton ansatz, keeping track of the non-conservative contributions, the latter of which, if desired,  can be removed from the final answer to obtain the purely conservative result.

The presence of the singular point in the mode amplitudes  means we need to distinguish two separate analytic extensions, which we label with a $\eta =\pm1$ superscript, e.g. $\beta_{lm}^{\eta}$, corresponding to respectively whether we continue by going above or below the singular point in the complex $a^\star$ domain. Notice these extensions are conjugate with each other.
The non-rational functions in the mode coefficients are very sensitive to the choice of branch, fortunately, they are presented in certain combinations that make it easy to track the effects of their contributions to one branch or the other. As one might expect,  such a  branch choice is irrelevant for extracting the  conservative part of the amplitude, whereas, for the dissipative piece, it comes with an extra sign in some terms of the mode expansion. Explicitly in  our mode example above, the two choices of branch lead to the $a^\star\gg1$ expressions
\begin{align}\label{eq:mode_final_example}
\beta_{2,-1}^{(6)\eta}(\gamma)=-\frac{\sqrt{\pi/5}}{967680}a^5\sin^{3}\gamma\Big[12017+96768\alpha-\eta32256(1+4\alpha)\cos\gamma+17775\cos(2\gamma)\Big]\,.
\end{align}
Here $\eta=\pm1$ keeps track of the prescription for doing the analytic continuation, whereas $\alpha=1$, keeps track of the contributions from the  digamma functions. One can check explicitly that removing absorption before analytic continuation (with the procedure outlined above), leads to the conservative amplitude which agrees with the above expression for   $\eta=0$, or equivalently, averaging the $\pm$ results, then making  irrelevant  the branch choice. This was expected since as we mentioned already, the two choices for the analytic extension are conjugate with each other, having effectively the same effect of averaging the mode \eqref{eq:modes} with its complex conjugate and then taking $a^\star\gg1$. This feature continues to be true for the $l=2,3$ mode coefficients for $i=6,7$. Furthermore, notice in general  that absorptive pieces of the amplitude  ($\eta\ne0$) will have contributions from both, the $\sqrt{1-a^{\star2}}$ function,  as well as from the digamma functions. 

In summary, we have learned that the dissipative contribution to the classical amplitude comes with a non-zero $\eta$ contribution, whereas the conservative result follows from averaging the $\eta=\pm1$ solutions (or equivalently setting $\eta=0$). 
In \cref{tab:Teukolskysolutions} we summarize the three sets of solutions labeled by $\eta=0,\pm1$, which match the gravitational Compton ansatz in \eqref{eq:ansatzspin}. Note that, as advertised in the previous section, the dissipative contributions are captured purely by operators proportional to $|a|$.   

\subsection{Matching to the Compton ansatz}\label{sec:mathcing_comtpon}
BHPT modes of the form  \eqref{eq:mode_final_example} can now be matched to the Compton modes computed from \eqref{eq:compton_modes}. For instance, for our $l,m=2,-1$ example, and at order $a^5$,  the explicit form of the Compton mode is given by 

\begin{equation}
    \begin{split}
      \beta_{2,-1}^{(6)\text{QFT}}(\gamma)  = & -\frac{\sqrt{\pi/5}}{967680}a^5\sin^{3}\gamma\Big[12017+17775\cos(2\gamma)+20160(4+3\cos(2\gamma))c_{2}^{(0)}+60480c_{3}^{(0)}\\
      & -10080(7+6\cos(2\gamma))c_{2}^{(1)}
      -30240c_{3}^{(1)}
 +120960\cos\gamma\,\big(c_{2}^{(2)}\cos\gamma-c_{4}^{(0+1+2)}\big)\Big]\,,
    \end{split}
\end{equation}
where we have used the notation $c_4^{(0+1+2)}=c_4^{(0)}-c_4^{(1)}+c_4^{(2)}$. Analogous expressions follow for the different $l,m$ modes, at different orders in spin.  Comparison to the BHPT modes, and requiring condition \eqref{eq:comparison} to be satisfying for all $l,m$,  results in linear systems of  equations for the Compton coefficients $c_j^{(i)}$. Up to $a^6$, the explicit solutions are shown in the last column of  \cref{tab:Teukolskysolutions}. 
\subsection*{Outgoing boundary conditions }
In \cref{sec:anomalous} we have identified the $\eta=0$ solution as the conservative (horizon independent) part of the amplitude. However, consistency checks that support this identification need to be done. A first 
check is given by solving the scattering problem assuming that the boundary condition on the black hole horizon is \emph{outgoing} rather than ingoing. Fortunately, the relevant analytic information for the asymptotic amplitudes for this problem can also be gleaned from the work of Mano, Suzuki and Tagasugi \cite{Mano:1996vt} (their solution $R^{\rm out}$). Upon investigating this boundary condition, which alters the appropriate reflection and incidence coefficients of \eqref{eq:phase_shift}, we find that the solutions \cref{tab:Teukolskysolutions} that match the higher spin Compton ansatz are identical, modulo $\eta\rightarrow-\eta$. This therefore shows that 
 the $\eta=0$ solution  is indeed insensitive to the boundary conditions at the BH horizon. 

\subsubsection*{Helicity reversing amplitude}
In an analogous way, BHPT solutions for  the helicity reversing amplitude \eqref{eq:g_deff} can be matched to the minimal coupling amplitude \eqref{eq:same_helicity_compton}. Running similar analysis, we find that for Teukolsky solutions with boundary conditions of purely ingoing waves at the BH horizon, the exponential  \eqref{eq:same_helicity_compton} does not receive any modification at up to the sixth order in spin. Remarkably, BHPT solutions in this case are independent of the branch choice  used for the analitic continuation from   $a^\star<1$ to $a^\star\gg1$. 

\subsubsection*{Full series in $a^{\star}$ }

As anticipated in the previous section, by allowing the coefficients $c_j^{(i)}$ of our ansatz to depend on the parameter $a$, one can alternatively obtain an exact matching to BHPT, see \cref{tab:Teukolskyexact}. This is a non-trivial match since all the angular dependence on the vector $\vec{a}$, which is carried by the spherical harmonics in BHPT, is still captured by the EFT operators $\{k_2\cdot a,k_3\cdot a,w\cdot a\} $ in the ansatz. In turn, this eliminates the necessity to include an additional operator $|a|$ in our expansion, at the same time exhibiting a resumation of the BHPT result where all branch cuts are explicit in the polygamma functions.

\begin{table}
\begin{tabular}{|c|c|}
\hline 
Spin & Kerr Solution\tabularnewline
\hline 
$a^{5}$ & $\begin{aligned}c_{2}^{(i)} & =0\,,\,\,\,i=0,1,2\\
c_{3}^{(0)} & =\frac{128}{45a^{\star4}}(1+3a^{\star2})\Re\left(\psi^{(0)}(2i\frac{a^{\star}}{\hat{\kappa}})\right)\\
c_{3}^{(1)} & =\frac{16}{45a^{\star4}}\left((4-3a^{\star2})\Re\left(\psi^{(0)}(i\frac{a^{\star}}{\hat{\kappa}})\right)+12(1+3a^{\star2})\Re\left(\psi^{(0)}(2i\frac{a^{\star}}{\hat{\kappa}})\right)\right)\\
c_{4}^{(0)} & =\frac{32(1+3a^{\star2})}{45a^{\star5}}i\left(\hat{\kappa}-4\Im\left(\psi^{(0)}(2i\frac{a^{\star}}{\hat{\kappa}})\right)\right)\\
c_{4}^{(1)} & =\frac{8}{45a^{\star5}}i\Big((8+9a^{\star2})\hat{\kappa}-2a^{\star}(4-3a^{\star2})\Im\left(\psi^{(0)}(i\frac{a^{\star}}{\hat{\kappa}})\right)
\\
&
-16a^{\star}(1+3a^{\star2})\Im\left(\psi^{(0)}(2i\frac{a^{\star}}{\hat{\kappa}})\right)\Big)\\
c_{4}^{(2)} & =\frac{4}{45a^{\star5}}i\Big((2+6a^{\star2}-3a^{\star4})\hat{\kappa}-2a^{\star}(4-3a^{\star2})\Im\left(\psi^{(0)}(i\frac{a^{\star}}{\hat{\kappa}})\right)\\
&
-2a^{\star}(2+3a^{\star2})\Im\left(\psi^{(0)}(2i\frac{a^{\star}}{\hat{\kappa}})\right)\Big)
\end{aligned}
$\tabularnewline
\hline 
%\hline 
% & \tabularnewline
%\hline 
\end{tabular}

\caption{Exact matching to spin operators, where coefficients are relaxed to functions of the spin norm "$a$". Here $a^5$ refers to quintic monomials in $\{k_2\cdot a,k_3\cdot a, w\cdot a\}$ but to all orders in the norm. In the large $a$ limit, they reduce to the coefficients of \cref{tab:Teukolskysolutions}.}

\label{tab:Teukolskyexact}
\end{table}

\subsection*{Near zone/Far zone splitting}\label{sec:zone_splitting}

It has been suggested recently by the authors of \cite{Ivanov:2022qqt} to consider the following factorization of the ratio of the linear perturbation coefficients 
entering in the phase shift \eqref{eq:phase_shift},\footnote{The explicit form for the functions entering this ratio can be found e.g.\ in \cite{Sasaki:2003xr} and in Appendix A in Part I \cite{Bautista:2021wfy}.}
\begin{equation}\label{eq:factorization}
\frac{B_{\ell m\omega}^{(\text{refl})}}{B_{\ell m\omega}^{(\text{inc})}}=\frac{1}{\omega^{2s}}{\color{blue}\underbrace{\frac{1+ie^{i\pi\nu}\frac{K_{-\nu-1}}{K_{\nu}}}{1+ie^{-i\pi\nu}\frac{\sin(\pi(\nu-s+i\epsilon))}{\sin(\pi(\nu+s-i\epsilon))}\frac{K_{-\nu-1}}{K_{\nu}}}}_{\text{near\,zone}}}{\color{red}\underbrace{\times\frac{A_{-}^{\nu}}{A_{+}^{\nu}}e^{i\epsilon(2\ln\epsilon-(1-\kappa))}}_{\text{far\,zone}}}.
\end{equation}
They indicate  that finite-size effects should be encapsulated in the first ``near zone'' factor, whereas the second factor contains non-linearity effects of the gravitational field. Using this separation and  keeping the  contributions coming from the ``near zone'' term only, the authors demonstrated in a gauge invariant way the vanishing of the static Love numbers for a Kerr BH. In the context of gravitational wave scattering treated in this work, it is natural then to wonder to what extent the ``far zone'' factor leads to a scattering amplitude which can, on the one hand,  be compared to the minimal coupling expressions of \cref{compton-spin}, and on the other hand, be used to fix the free coefficients of the Compton ansatz for the higher spin  amplitude \eqref{eq:ansatzspin}.

Running an analysis  analogous to the one used in \cref{Sec:PWScattering}, up to order $\epsilon^6$ in \eqref{eq:beta_modes}, ``far zone'' solutions produce  scattering amplitudes that are purely polynomial  in $a^\star$,\footnote{This is expected as polygamma contributions are present only in the $K_\nu$ and $K_{-\nu-1}$ functions in \eqref{eq:factorization}. } and have therefore  a unique (trivial) $a^\star\gg1$ extension. Up to order $a^3$,  the  ``far zone'' solutions produce an amplitude that matches precisely up to  the third order in the $a^\star$-expansion of the exponential in  \eqref{eq:ansatzspin}. 
At the fourth order in spin, the  ``far zone'' solution does not match exactly the exponential amplitude; however, the resulting amplitude can still be accommodated in the Compton ansatz \eqref{eq:ansatzspin}, with   contact terms  modifications summarized in  \cref{tab:far_zone}. This solution  breaks explicitly the spin-shift symmetry.

In order to match the ``far zone'' solution at the fifth order in spin, we find the ansatz  \eqref{eq:ansatzspin} needs to be enlarged by the non-contact terms 
\begin{equation}\label{eq:extra}
\begin{split}
    \Delta P_\xi =& e_{1}^{(0)}\frac{(w{\cdot}a)^{5}}{\xi^{2}}+e_{1}^{(1)}\frac{(w{\cdot}a)^{5}(k_{2}{\cdot}a)(-k_{3}{\cdot}a)}{\xi}\\
    &+(w{\cdot}a-k_{2}{\cdot}a)(w{\cdot}a+k_{3}{\cdot}a)w{\cdot}a\left(e_{1}^{(2)}(k_{2}{\cdot}a)(-k_{3}{\cdot}a)-e_{1}^{(0)}(w{\cdot}a)^{2}\right)\,,
\end{split}
\end{equation}
with the values of the coefficients given in \cref{tab:far_zone}. Interestingly, it is easily checked that these extra terms preserve the t-channel residue but modify the s- and u- channel exchanges. This signals that the far-zone amplitude proposed in \eqref{eq:factorization}    produces the effect of additional massive states (different to Kerr) propagating in the $s$ channel. It would be interesting to explore this modification further \footnote{From the QFT point of view, one can expect that they correspond to 3-pt amplitudes of a spin $S$ particle decaying into a spin $S'$, where both $S,S' \to \infty$.}.

\begin{table}
\begin{centering}
\begin{tabular}{|c|c|}
\hline 
Spin & ``Far zone'' solutions\tabularnewline
\hline 
\hline 
$a^{4}$ & $c_{1}^{(0)}=-\frac{189056}{103041},\,\,c_{1}^{(1)}=-\frac{86044}{34347},\,\,,c_{1}^{(2)}=-\frac{10402}{34347},\,\,$\tabularnewline
\hline 
\hline 
$a^{5}$ & $\begin{aligned}c_{2}^{(0)} & =\frac{114208}{148837},\,\,c_{2}^{(1)}=\frac{1130912}{744185},\,\,c_{2}^{(2)}=\frac{435488}{2232555}\\
c_{3}^{(0)} & =\frac{286064608}{1157665635},\,\,c_{3}^{(1)}=-\frac{2531080196}{15049653255},\,\,c_{3}^{(2)}=-\frac{745559744}{5016551085}\\
c_{4}^{(i)} & =0,\,\,i=0,1,2,\,\,e_{1}^{(0)}=\frac{64}{195}\,\,,e_{1}^{(1)}=\frac{32}{117}\,\,,e_{1}^{(2)}=\frac{8}{195}\,\,
\end{aligned}
$\tabularnewline
\hline 
\end{tabular}
\caption{Spin 4 and 5, ``Far zone'' Teukolsky solutions}
\label{tab:far_zone}
\par\end{centering}
\end{table}

\section{Discussion}\label{sec:discussion}
In this paper, we showed how to extract a tree-level gravitational Compton amplitude  from solutions of the Teukolsky equation, up to the sixth order in the BH's spin, for both the helicity preserving ($f$) and helicity reversing ($g$) cases. We showed by explicit computation that up to the considered order in spin,  there is a contribution  ($\eta=0$) to the helicity preserving amplitude  that does not depend on the boundary conditions at the horizon, which we therefore  identify  with the \textit{conservative} part of the amplitude. On the other hand, for the same amplitude,   we have seen that BH horizon dissipative effects can  easily be encapsulated by operators in the gravitational Compton amplitude proportional to $|a|$, therefore enlarging the usual 3-dimensional BH spin basis. Investigation of the effect of this spin basis enlargement  on the spin supplementary condition needs to be further investigated. 
For the  helicity reversing scenario, a unique conservative (only real contributions) amplitude has been extracted from low energy  Teukolsky solutions  up to sixth order in BH spin, whose 
whose spin structure coincides exactly with the truncation of the  exponential $e^{k_2\cdot a+k_3\cdot a}$ up to sixth order in $a$.

Although the conservative solutions have a unique identification, dissipative effects ($\eta=\pm1$) are sensitive to the choice of branch  made when doing the continuation of Teukolsky solutions from $a^\star<1$ to $a^\star\gg1$. This is a consequence of erasing the BH horizon by the analytic continuation, therefore obscuring the uniqueness of the interpretation of the physical behavior of  gravitational waves at the BH horizon. Further investigation regarding this non-uniqueness of the imaginary contributions to the amplitude  is left for future work. 
Notice however, the issues with  identifying a unique dissipative contribution to the Compton amplitude have origin in the analytic continuation. This issue could be avoided by looking for low energy amplitudes from the solution to the Teukolsky equation,  but keeping  all orders in $a/GM$, 
%therefore proving the actual Kerr BH, 
as we have done in  
\cref{tab:Teukolskyexact}. Further research   in this direction is left for future work.

It is important to note, however, that the complications of analytic continuation and related issues affect our extraction of the tree-level Compton amplitude starting only from the fifth order in spin.  Up to fourth order in spin, we have found that the unambiguously unique tree-level part of the Teukolsky amplitude coincides precisely with the previously conjectured exponential form of the classical Compton amplitude \cite{Guevara:2018wpp,Chung:2018kqs,Aoude:2020onz} arising from a classical or heavy-particle limit from the ``minimally coupled'' Compton amplitude originally presented in \cite{Arkani-Hamed:2017jhn}.  This provides a significantly more complete justification that the results for the 2PM (or next-to-leading-order PN) conservative dynamics of two-BH systems presented in \cite{Guevara:2018wpp,Chen:2021kxt,Bern:2022kto,Aoude:2022trd,Levi:2022rrq} do indeed fully correspond to predictions of GR through fourth order in spin.  Concerning the recent predictions at the fifth order in spin (and beyond) for our Compton amplitude,  it is  left for   future work to further use it in the  two-BH problem at higher orders in spins and with generic spin orientations.

Finally, it would be interesting to study  the double copy structure of the gravitational Compton amplitude obtained from Teukolsky solutions. The ansatz for the higher spin Compton amplitude \eqref{eq:ansatzspin} was built by factorizing the scalar amplitude, therefore intrinsically introducing a double copy structure in the amplitude. The single copy amplitude to study in the context of wave perturbations will be spin $s=-1$ perturbations of  $\sqrt{\text{Kerr}}$ \cite{Arkani-Hamed:2019ymq}. Since in  the static configuration,  $\sqrt{\text{Kerr}}$ is  obtained from the Kerr-Newman solution in the $GM\to0$ limit, a wave equation analog to the Teukolsky equation may be derivable from electromagnetic perturbations on a Kerr-Newman background, setting  $GM\to 0$ at the end of the computation. Exploration of this idea is left for future work.

\appendix

\acknowledgments
We thank  Francesco Alessio, Stefano de Angelis, Zvi Bern, Lucille Cangemi, Marco Chiodaroli, Gustav Jakobsen, Henrik Johansson, Dimitris Kosmopoulos, David Kosower, Andr\'es Luna, Gustav Mogull, Julio Parra-Martinez, Alexander Ochirov, Donal O'Connell, Jan Plefka, Radu Roiban, M.\ V.\ S.\ Saketh, Matteo Sergola, Chia-Hsien Shen, Nils Siemonsen, Jan Steinhoff, Fei Teng, and Mao Zeng for useful discussions. We are grateful to  Rafael Aoude, Kays Haddad and  Andreas Helset for agreeing to exchange preliminary drafts of our works, including \cite{Aoude:2022trd}. A.G. and J.V. are grateful for hospitality at KITP Santa Barbara, during the program `High-Precision Gravitational Waves'. A.G. is supported by a Junior Fellowship at the Harvard Society of Fellows, as well as by the DOE grant de-sc/0007870. The work of Y.F.B. has been supported in part by the European Research Council under Advanced Investigator Grant ERC–AdG–885414.  Research at Perimeter Institute is supported by the Government of Canada through the Department of Innovation, Science and Economic Development Canada and by the Province of Ontario through the
Ministry of Research, Innovation and Science.  This publication has emanated from research supported in part by a Grant from Science Foundation Ireland under Grant number 21/PATH-S/9610.

\appendix

\section{Chiral Conjugate Amplitude}\label{ap:chiral}

In this appendix we derive the following relation between conjugate amplitudes, defined in \eqref{eq:chiral-to-antichiral}:
\begin{equation}\label{eq:ap1}
    A_n^{\textrm{antichir.}} = e^{2K\cdot \tilde{\mathbb{W}}} A_n^{\textrm{chir.}|_{\mathbb{W}\to\tilde{\mathbb{W}}}} +\mathcal{O}(\hbar)
\end{equation}
Recall $\mathbb{W}$ is an operator acting on  spin-$S$ representations. More formally, we should use the notation $\mathbb{W}_S$, but we chose to omit the spin-subscript in order to simplify the notation. 
As both amplitudes are functions of operators $q_i\cdot\mathbb{W}$, for a certain set of momentum transfer variables $\{q_i\}\sim \hbar$, it suffices to show the following
\begin{equation}\label{eq:toprove}
   \langle \varepsilon_n | \prod_{i=1}^{j} q_i \cdot \mathbb{W} |\varepsilon_1\rangle = [\varepsilon_n |e^{2K\cdot \tilde{\mathbb{W}}} \prod_{i=1}^{j} q_i \cdot \tilde{\mathbb{W}} |\varepsilon_1]  +\mathcal{O}(\hbar) \,.
\end{equation}
We identify $K=p_n-p_1$ as one of the momentum transfer variables, e.g. $K=q_{j+1}$. In order to proceed we introduce the spin-1/2 chiral representation of the Pauli-Lubanski operator, denoted by $\frak{a}^\mu\equiv \mathbb{W}_{\frac{1}{2}}$. According to eqs. \eqref{eq:BandS}, \eqref{bprop} we have
\begin{equation}\label{eq:aspin12}
    \frak{J}_{\mu \nu}= \frac{i}{4}(\sigma_\mu \tilde{\sigma}_\nu -\sigma_\nu \tilde{\sigma}_\mu) \Longrightarrow \frak{a}^\mu =\frac{u^{\nu} \frak{J}_{\nu \mu}}{iM}=\frac{1}{4M} (U \tilde{\sigma}^{\mu} - \sigma^\mu \tilde{U})\,.
\end{equation}
where we are using $u^\mu=\frac{p_1^\mu}{M}$, and $U=u_\mu \sigma^\mu, \tilde{U}=u_\mu \tilde{\sigma}^\mu$. Defining the antichiral conjugate $\tilde{\frak{a}}^\mu$ we note that 
\begin{equation}\label{eq:auisua}
    \frak{a}^{\mu} U = U \tilde{\frak{a}}^\mu \,.
\end{equation}
which is the infinitesimal version of $U=e^{-q\cdot \frak{a}} U  e^{q\cdot \frak{\tilde{a}}}$, manifesting the fact that $\frak{a}^\mu$ is a little group generator. Furthermore, the operators $\frak{a}^\mu$ are a covariant version of the Pauli matrices and hence lead to the algebra.

\begin{equation}
    M \frak{a}^\mu  \frak{a}^\nu = M^{-1}(\eta^{\mu \nu} - u^\mu u^\nu)\mathbb{I} + S^{\mu \nu}\,. 
\end{equation}
Now we introduce the notation $\frak{a}_i:= q_i \cdot \frak{a}$. Note that according to the scaling rules $\frak{a}_i\sim \hbar^0$, i.e. they are classical variables. Nevertheless, their product satisfies
\begin{equation}\label{eq:aaprod}
    M \frak{a}_i \frak{a}_j = M^{-1}\left(q_i \cdot q_j- (q_i\cdot u) (q_j\cdot u)\right)\mathbb{I} + q_i\cdot S\cdot q_j \,,
\end{equation}
which scales as $\mathcal{O}(\hbar)$ and hence we can drop it as a quantum contribution.\footnote{In certain cases the RHS of \eqref{eq:aaprod} actually vanishes identically, for example for the combination $q_i=q_j=w-k_2$ appearing in $A_4$ for low spins.} This reflects the absence of quadrupole operator for spin-1/2 states. In the following, we will also need the identity 
\begin{align}
   U\tilde{q}_i &= u_\mu q_\nu \sigma^\mu \tilde{\sigma}^{\nu} =  2M\frak{a}_i + (u\cdot q_i)\mathbb{I} \\
    \tilde{q}_i U &= 2M\tilde{\frak{a}}_i + (u\cdot q_i)\mathbb{I} \,\, , \label{eq:qiUm}
\end{align}
which easily follows from \eqref{eq:aspin12} together with $\sigma^{(\mu} \tilde{\sigma}^{\nu)}=\eta^{\mu \nu}\mathbb{I}$.

We are now ready to provide a derivation of \eqref{eq:toprove}. From the construction of the tensor product states \eqref{basi1}, we know that the spin-$S$ generator $\mathbb{W}^\mu $ in \eqref{eq:toprove} is the direct sum of spin-1/2 generators. Then, we have

\begin{equation}\label{eq:aisum}
    \mathbb{W} _i := q_i\cdot  \mathbb{W} = \frak{a}_i\otimes \mathbb{I}^{\otimes 2S-1}+\mathbb{I}\otimes \frak{a}_i\otimes \mathbb{I}^{\otimes 2S-2} + \ldots\,.
\end{equation}
Inserting this in the LHS of \eqref{eq:toprove} leads to

\begin{equation}\label{eq:toprove2}
   \langle \varepsilon_n | \prod_{i=1}^{j}  \mathbb{W}_i |\varepsilon_1\rangle = j!\binom{2S}{j}  \langle \varepsilon_n | \frak{a}_1\otimes \frak{a}_2\otimes \ldots \otimes \frak{a}_j\otimes \mathbb{I}^{\otimes 2S-j}  |\varepsilon_1\rangle  +\mathcal{O}(\hbar) \,.
\end{equation}
(we used that states $|\varepsilon_1\rangle, |\varepsilon_n\rangle  $ are constructed as symmetrized tensor products, together with the nilpotent condition $\frak{a}_i \frak{a}_j\sim \hbar$, which forces the operators $\frak{a}_i$ to occupy one slot each in the tensor product). Note that this formula imposes $j\leq 2S$, which shows that a spin-$S$ particle does not lead to classical contributions beyond $a^{2S}$. 

To change the basis of states we use
\begin{align}
     \langle \varepsilon_n | &= [\varepsilon_n| (\tilde{U}+\frac{1}{M}\tilde{K})^{\otimes 2S} \\
     |\varepsilon_1\rangle &= U^{\otimes 2S} |\varepsilon_1 ]
\end{align}
because $p^\mu_n=Mu^\mu + K^{\mu}$. Further dropping $\tilde{K}\frak{a}_i\sim \hbar$, and using \eqref{eq:auisua} and \eqref{eq:qiUm} for $\tilde{q}_i=\tilde{K}$ we obtain

\begin{align}\label{eq:toprove3}
   \langle \varepsilon_n | \prod_{i=1}^{j}  \mathbb{W}_i |\varepsilon_1\rangle &= j! \binom{2S}{j}  [\varepsilon_n | (\tilde{U}\frak{a}_1 U )\otimes (\tilde{U} \frak{a}_2 U )\otimes \ldots \otimes (\tilde{U} \frak{a}_j U )\otimes (\mathbb{I}+\frac{1}{M}\tilde{K}U)^{\otimes 2S-j}  |\varepsilon_1] +\mathcal{O}(\hbar) \, \nonumber \\
   &= j!  \binom{2S}{j}  [\varepsilon_n | \tilde{\frak{a}}_1 \otimes \tilde{\frak{a}}_2 \otimes \ldots \otimes \tilde{\frak{a}}_j \otimes (\mathbb{I}+2\tilde{\frak{a}}\cdot K)^{\otimes 2S-j}  |\varepsilon_1] +\mathcal{O}(\hbar) \, \nonumber \\
   &= \sum_{l=0}^{2S-j} j! \binom{2S}{j} \binom{2S-j}{l}  [\varepsilon_n | \tilde{\frak{a}}_1 \otimes \tilde{\frak{a}}_2 \otimes \ldots \otimes \tilde{\frak{a}}_j \otimes (2\tilde{\frak{a}}\cdot K)^{\otimes l}\otimes \mathbb{I}^{\otimes 2S-j-l}  |\varepsilon_1] +\mathcal{O}(\hbar) \, \nonumber \\
   &= \sum_{l=0}^{\infty}\frac{j!}{(j+l)!} \underbrace{\binom{2S}{j+l}^{-1}\binom{2S}{j} \binom{2S-j}{l}}_{\binom{l+j}{j}} [\varepsilon_n | (2 K\cdot \tilde{ \mathbb{W}})^l \prod_{i=1}^{j} \tilde{ \mathbb{W}}_i  |\varepsilon_1] +\mathcal{O}(\hbar) \, .
\end{align}
In the last line, we have applied again the identity \eqref{eq:toprove2}, this time for the antichiral state basis. The sum can be extended to the infinite series since, as we argued, it leads to quantum corrections for $l+j>2S$. Thus we obtain a formal exponentiation, which ends the proof of \eqref{eq:ap1}.

\section{Classical Crossing Symmetry}\label{ap:bose}
In this Appendix, we derive the classical constraint \eqref{eq:a4sbox} arising from Crossing symmetry. Using the definition \eqref{eq:xidef} we can write the scalar component of the Compton amplitude \eqref{eq:ansatzspin},  simply as 
\begin{equation}\label{eq:a40xi}
    A_4^0= 32\pi G M^2\times \frac{(\epsilon_2\cdot u)^2(\tilde{\epsilon}_3\cdot u)^2}{\xi}\,.
\end{equation}
Notice that  because of our gauge fixing \eqref{eq:chkgauge}, the product of polarization vectors  also generates an overall pole in $t$. 

We now introduce the chiral conjugate amplitude $\hat{A}_4^S$ by flipping the helicity of the two gravitons: We turn $k_2$ into negative helicity and $k_3$ into positive helicity. It reads

\begin{equation}\label{eq:ahat4def}
    \hat{A}_4^S  = [\varepsilon_4| \hat{A}^{\textrm{antichir.}}_4|\varepsilon_1]
\end{equation}
Due to parity conservation, we note that $\hat{A}_4^S$ is obtained from $A_4^S=\langle \varepsilon_4|A_4^{\textrm{chir.}}|\varepsilon_1\rangle$ by formally swapping $\sigma^\mu \leftrightarrow \tilde{\sigma}^\mu$, including the exchange of angle and square brackets. If $A_4^{\textrm{chir.}}$ is expanded in spin operators $a^\mu$, chiral conjugation implies that we must take $a^\mu \to -\tilde{a}^\mu$ (the minus sign coming from \eqref{spinvecac}, from which $\tilde{\mathbb{J}}^{\mu\nu}$ is now anti self-dual\footnote{Chiral conjugation can also be formulated as a PT transformation, for which the spin pseudovector will flip sign.}). In other words, from the form \eqref{eq:fform} we have

\begin{equation}\label{eq:tildew}
    \hat{A}^{\textrm{antichir.}}_4 = \hat{A}_4^0 \times P_\xi (-k_2 \cdot \tilde{a}, -k_3\cdot \tilde{a}, - \tilde{w}\cdot\tilde{a}) \,,\quad \tilde{w}^\mu=\frac{u\cdot k_2}{u\cdot \tilde{\epsilon_2}}  \tilde{\epsilon}_2^\mu \,.
\end{equation}
Crossing symmetry is the fact that $A_4^S$ can also be obtained from $\hat{A}_4^S$ by swapping particle labels $2 \leftrightarrow 3$, which physically swaps the channels $s$ and $u$ (even though at this point we are working with the analytically continued 
both-incoming amplitude). Using definitions \eqref{eq:chiral-to-antichiral} and \eqref{eq:ahat4def}, this implies that   $A^{\textrm{antichir.}}_4$ can be obtained from $\hat{A}^{\textrm{antichir.}}_4$ via the same procedure. Thus, 

\begin{equation}\label{eq:a4antic}
   A^{\textrm{antichir.}}_4 = A_4^0 \times P_\xi (-k_3 \cdot \tilde{a}, -k_2 \cdot \tilde{a},
    - w'\cdot\tilde{a}) \,,\quad w'^\mu=\frac{u\cdot k_3}{u\cdot \tilde{\epsilon_3}}  \tilde{\epsilon}_3^\mu \,,
\end{equation}
where we used that the ratio $\xi$ is invariant under the exchange. We can now finally use \eqref{eq:chiralityflip} to compare \eqref{eq:a4antic} with \eqref{eq:fform} as $\hbar \to 0$. We obtain

\begin{equation}
  P_\xi (-k_3 \cdot \tilde{a}, -k_2 \cdot \tilde{a},w\cdot \tilde{a}) = e^{2(k_2+k_3)\cdot \tilde{a}} P_\xi (k_2\cdot \tilde{a},k_3 \cdot \tilde{a}, w\cdot \tilde{a}) +\mathcal{O}(\hbar) \,, 
\end{equation}
where we used that in the classical limit $w'=\frac{u\cdot k_3}{u\cdot \tilde{\epsilon_3}}  \tilde{\epsilon}_3 \to  -\frac{u\cdot k_2}{u\cdot \epsilon_2}  \epsilon_2 =- w$. Another way to phrase this result is, from \eqref{eq:ansclas}, given by \eqref{eq:a4sbox}.

\section{Spinning Spherical Harmonics as EFT spin operators}\label{sec:spinningspheroidalharmonics}

In the main text we have constructed an ansatz for the Compton amplitude in terms of the spin operators $\{k_2\cdot a,k_3\cdot a,w\cdot a\}$. In contrast, the results from BHPT are given in terms of spinning spherical harmonics $\,_{h}Y_{lm}(\theta,\phi)$. The matching can be done for instance by projection of the ansatz into harmonics, c.f. \eqref{eq:compton_modes}. In this Appendix we provide an alternative: We outline a direct construction to rewrite each combination of $\,_{h}Y_{lm}(\theta,\phi)$ explicitly in terms of spin operators, without the need of performing the projection integrals.

\subsection{Harmonics from Massive Spinors}

First, we provide a representation of the spinning spherical
harmonics \cite{Goldberg:1966uu} using the spinor variables of \cref{compton-spin}. This representation
allow us to translate the Compton amplitude written in the latter
variables as a sum of partial waves.

Spherical harmonics are irreducible representations of the massive
little group $su(2)$, see Appendix III in  \cite{Guevara:2021yud} for a more detailed discussion.
This requires specifying a time direction, breaking Lorentz symmetry
into $su(2)$, which here we take as
\begin{equation}
   U=\frac{1}{M}\sigma_\mu p_1^\mu =|1^{a}\rangle[1_{a}|= |1^+\rangle [1^-| - |1^-\rangle [1^+|\,,
\end{equation}
To define spherical harmonics we construct functions on the celestial
sphere $\mathbb{CP}^{1}$ on which $su(2)$ acts via rotations. Furthermore,
we refer to the harmonics as spinning if the functions carry helicity
weight $h$ under the massless little group SO$(2)$. Recall we have
associated the null direction with the massless momenta of the outgoing
gravitational wave, namely

\begin{equation}
k_{3}=|\lambda\rangle[\lambda|\,\,,
\end{equation}
To introduce the projective variables defined on the sphere we remove
an overall energy scale: Define

\begin{equation}
\hat{k}:=\frac{k_{3}}{2u \cdot k_{3}}=\frac{|\lambda\rangle[\lambda|}{\langle\lambda|U|\lambda]}\,,
\end{equation}
which indeed is invariant under projective rescalings of $|\lambda\rangle,[\lambda|$, 
\begin{equation}
|  \lambda\rangle\to t|\lambda\rangle\,,|\lambda]\to\tilde{t}|\lambda]\,,  
\end{equation}
and hence carries no helicity weight. We then introduce, extending
the discussion of e.g. \cite{Guevara:2021yud}, 
\begin{align}
\,_{h}Y_{lm}(\theta,\phi) & =\frac{1}{\langle\lambda|U|\lambda]^{l}}\underbrace{\langle\lambda1^{(a_{1}}\rangle\cdots\langle\lambda1^{a_{l-h}}\rangle}_{l-h}\underbrace{[\lambda1^{a_{l-h+1}}]\cdots[\lambda1^{a_{2l)}}]}_{l+h}\,\,,l\geq h\,,
\end{align}
such that
\begin{equation}
    (a_1,\ldots,a_{2l})=(\underbrace{+,\ldots,+}_{l+m},\underbrace{-,\ldots,-}_{l-m})\,.
\end{equation}
This transforms as $\,_{h}Y_{lm}\to(t/\tilde{t})^{h}\,_{h}Y_{lm}$
and hence carries helicity, or spin-weight, $h$. We will make the $su(2)$
indices $a_{i}=1,2$, together with its complete symmetrization, implicit
in the following. It is easy to show that this agrees with the definition
of the harmonics given in \cite{Goldberg:1966uu} by introducing stereographic coordinates
in $\mathbb{CP}^{1}$, via $|\lambda\rangle=(1\,\,z)$ and $|\lambda]=(1\,\,\,\bar{z})$. 

For our purposes, we will parametrize the outgoing massless momenta
in terms of the two angles of the celestial sphere, 
\begin{equation}
  k_{3}^{\mu}=\omega(1,\hat{n}(\theta,\phi))  \,,
\end{equation}
for the energy scale $\omega$, and a unit vector $\hat{n}$. Then,
the spinors give the natural embedding from the Bloch sphere 

\begin{align}
|\lambda\rangle & =\sqrt{2\omega}\left(-e^{-i\phi/2}\sin(\theta/2)|1^+\rangle +e^{i\phi/2}\cos(\theta/2)|1^-\rangle \right)\,,\\{}
[\lambda| & =\sqrt{2\omega}\left(-e^{i\phi/2}\sin(\theta/2)[1^-|+e^{-i\phi/2}\cos(\theta/2)[1^+|\right)\,,
\end{align}
which correspond to the spinors introduced in \eqref{eq:spinhel}. In these coordinates, the incoming $k_2^\mu$ is obtained by setting $\theta=0$ and hence 

\begin{equation}\label{eq:k2sx}
    \hat{k}_2 = \frac{k_2}{2u\cdot k_2}=  | 1^-\rangle  [1^+|\,.
\end{equation}
In this notation we find that the main text t-channel is
\begin{equation}\label{eq:xisda}
    \xi = \frac{1}{\hat{k}_2 \cdot \hat{k}} = \frac{\langle \lambda| U|\lambda]}{\langle \lambda -\rangle [\lambda +]} = \frac{\langle \lambda +\rangle [\lambda -]}{\langle \lambda -\rangle [\lambda +]} - 1\,.
\end{equation}
More generally, for any function of $|\lambda\rangle, |\lambda ] \in \mathbb{CP}^1$ we can introduce the helicity $h$ as the weight of $|\lambda\rangle$ minus the weight of $|\lambda ]$ (here a spinor is weight $1/2$). We can also introduce its `azimuthal weight' $m$ as the weight in $|\pm\rangle,|\pm ]$. For instance \eqref{eq:k2sx} has azimuthal weight $m=0$ (and $h=0$), which means it is aligned with the z axis. One can then think of the polarizations \eqref{eq:chkgauge} 

\begin{equation}\label{eq:sxw}
    \epsilon_2^- \cdot u = \frac{[\lambda|U|-\rangle}{[\lambda +]}= \frac{[\lambda -]}{[\lambda +]}\,,\quad ,   \epsilon_3^+ \cdot u = \frac{[\lambda|U|-\rangle}{\langle \lambda -\rangle}= \frac{[\lambda -]}{\langle \lambda -\rangle}\,,
\end{equation}
as having, respectively, $(h,m)=(0,-1)$ and $(h,m)=(1,0)$. Thus, in aligning $k_2$ with the $z$ axis we have effectively turned its corresponding helicity into azimuthal weight. Indeed, using this we see that the scalar amplitude

\begin{equation}
    A_4^0= 32\pi G M^2\times \frac{(\epsilon_2\cdot u)^2(\tilde{\epsilon}_3\cdot u)^2}{\xi}\,,
\end{equation}
has $(h,m)=(2,-2)$ (it follows from \eqref{eq:xisda} that $\xi$ has weight $(0,0)$). As a consequence it can be expanded into all harmonics consistent with those quantum numbers, in the form
\begin{equation}\label{eq:somea04ex}
    A_4^0= \sum^{\ell_{\textrm{max}}=\infty}_{\ell\geq 2}  \,_{2}Y_{\ell,-2}(\theta,\phi) c_\ell = \,_{2}Y_{2,-2}(\theta,\phi) f(\xi)\,.
\end{equation}
where $c_{\ell}$ are numerical constants (independent of $\xi$). The upper limit $\ell_{\textrm{max}}$ can be obtained by expanding in powers of $\langle\lambda|U|\lambda]$, in this case the series does not truncate. The second equality, involving a scalar function $f(\xi)$, follows from a simple lemma: Any function with $(h,m)=0$ in $\mathbb{CP}^1$ can be written solely as a function of \eqref{eq:xisda}. Explicitly in this case $f(\xi)=32GM^2 \xi$.

\subsection{Relation to Spin Operators}\label{sec:apc2}

We are now in position to rewrite spherical harmonics in terms of the spin operators of section 3.3. The main ingredient is the relation
\begin{equation}
    \frac{u\cdot F_2^- \cdot a}{\epsilon_2\cdot u} = w\cdot a- k_2\cdot a\,,
\end{equation}
where we have emphasized the helicity $-2$ of $k_2$. We should temporarily also consider the other helicity configuration as well. This is simply the chiral conjugate of the above. Using \eqref{eq:tildew} this is

\begin{equation}
    \frac{u\cdot F_2^+ \cdot a}{\epsilon_2\cdot u} = \tilde{w}\cdot a- k_2\cdot a\,.
\end{equation}
Using \eqref{eq:sxw} we can rewrite this as 
\begin{equation}\label{eq:restf}
    \frac{[\lambda +]}{[\lambda -]} a_{++} = (w-k_2)\cdot a \,,\quad   \frac{\langle \lambda -\rangle}{\langle\lambda +\rangle} a_{--} = (\tilde{w}-k_2)\cdot a\,,
\end{equation}
where $a_{\pm \pm}=\omega [\pm|a|\pm\rangle$. Furthermore
\begin{equation}
    k_2\cdot a = a_{+-}\,.
\end{equation}
Thus we have identified three operators, each with either $m=-1,0,1$ and $h=0$.\footnote{They span a $SL(2,\mathbb{R})$ algebra $L_{\frac{m+n}{2}}=a_{m,n}$.} (The relation to the basis $\{k_2\cdot a, k_3 \cdot a, w\cdot a\}$ is obtained from \eqref{eq:wpluswtilde}.) Products of them span the spherical harmonics of the previous section, modulo functions $f(\xi)$. For instance, for $m\geq -2$, at order $(a\omega)^j$, we can use the ansatz

\begin{equation}
     (a\omega)^j \,_{2}Y_{\ell,m} = f_\ell (\xi,\gamma) (\epsilon_2^- \cdot u)^2 (\epsilon_3^+\cdot u)^2 (\tilde{w}\cdot a - k_2\cdot a)^{m+2} (k_2\cdot a)^{j-m-2}\,,
\end{equation}
and easily solve for the function $f_\ell (\xi,\gamma)$ at each $\ell$. Let us explain the dependence on $\gamma$, defined by \eqref{eq:axayaz}. Since in our frame $k_3=k_3 (\theta,\phi)$, we can perform a $z-$rotation to align $a^\mu$ to the $x-z$ plane, setting $\phi'=0$ in by \eqref{eq:axayaz}. This yields
\begin{equation}
    a_{++}=-a_{--}=\frac{a \omega}{2} \sin \gamma\,,\quad a_{+-}=\frac{a \omega}{2} \cos \gamma\,.
\end{equation}
Now, note that the coefficients $\beta_{\ell m}$ \eqref{eq:beta_modes} in the harmonic expansion also carry factors of $\cos\gamma$ and $\sin \gamma$. Factors of $\cos\gamma$ can be exchanged by the operators $k_2\cdot a/|a|\omega$ introduced in section 3. Furthermore, factors of $\sin^2\gamma$ can be exchanged by the $m=0$ combination
\begin{equation}
  4\frac{ [(w-k_2)\cdot a] [(\tilde{w}-k_2)\cdot a]}{(|a|\omega)^2 }= -\frac{1}{\xi+1} \sin^2 \gamma \,.
\end{equation}
By rewriting all the harmonics in the series \eqref{eq:fylmex} in terms of spin operators we expect spurious poles to cancel at each order in $a\omega$. Note that \eqref{eq:a2identity} is equivalent to the identity 
\begin{equation}
\frac{a^2\omega^2}{4}=a_{++}a_{--} - a_{+-}^2\,,
\end{equation}
which can be used to eliminate the operator $|a|$. As our final example, consider the exponential part of the Compton amplitude. It can be written as a combination of $m=-1,0,1$ terms:

\begin{equation}\label{eq:expdecp}
    \langle A_4^S\rangle =\,_{2}Y_{2,-2}(\theta,\phi) \times 32GM^2 \xi \left(e^{\frac{\xi-1}{\xi}(k_2-w)\cdot a - \frac{\xi+1}{\xi} (k_2 - \tilde{w})\cdot a + \frac{2}{\xi} k_2\cdot a 
 }+\ldots \right)\,.
\end{equation}
Let us consider the polar setup $\gamma=0$, meaning $a_{++}=a_{--}=0$. In this case the first two terms in the exponent vanish due to \eqref{eq:restf}, as well as the contact terms given by $(\ldots)$. By expanding in powers of $a$ we are left with the series

\begin{equation}
     \langle A_4^S\rangle_{\gamma\to 0}=\,_{2}Y_{2,-2}(\theta,\phi) \times 32GM^2 \xi  \sum_k \frac{1}{k!}\left(\frac{2}{\xi} k_2\cdot a \right)^k\,.
\end{equation}
Since $k_2 \cdot a= a\omega/2$, we see that at order $(a\omega)^k$ we obtain a divergence $\xi^{1-k} \sim \frac{1}{\langle \lambda |U|\lambda ]^{k-1}}$. Together with the fact that $k_2 \cdot a$ has $m=0$, this means that at this order the amplitude can be expanded in the form \eqref{eq:somea04ex} with $\ell_{\textrm{max}}=k-1$. We have explicitly checked this in the polar BHPT computation, in contrast with the non-polar case in which the series does not truncate.

\section{Spin-Shift Symmetry}\label{sec:shift_symmetry}
Inspired by lower spin results, recently, it has been proposed by the authors of \cite{Bern:2022kto,Aoude:2022trd},  the higher-spin ($\mathcal{O}(a^{l>4})$) 2PM  amplitude for the Eikonal scattering of two Kerr BHs, 
should respect certain symmetry associated with the shift of the spins of the  bodies by an amount proportional to the momentum transfer of the massive $2\to2$  scattering process.

The aforementioned 2PM amplitude can be computed for instance  from the unitary gluing of two gravitational Compton amplitudes $\langle A_4\rangle$. Up to the fourth order in spin, only the opposite helicity configuration for the Compton amplitude contributes to the final result  \cite{Chen:2021kxt}. In the aligned spin scenario, it was checked by explicit computation in \cite{Guevara:2018wpp}, the same helicity Compton amplitude does not contribute to the 2PM scattering angle.

\begin{table}
\begin{center}
\begin{tabular}{|c|c|c|c|}
\hline 
Spin & Shift-Sym. & Free Coeffs. & Relation to  \cite{Aoude:2022trd}
\tabularnewline
\hline 
\hline 
$a^{4}$ & $\begin{aligned}c_{1}^{(i)} & =0,\,i=1,2\end{aligned}
$ & $c_{1}^{(0)}$ & $c_{1}^{(0)}=-\frac{d_{0}^{(4)}}{4!}$\tabularnewline
\hline 
\hline 
$a^{5}$ & $\begin{aligned}c_{j}^{(i)} & =0,\,\,i=1,2,\,\,j=2,3\\
c_{3}^{(0)} & =\frac{4}{15},\,\,c_{4}^{(i)}=0\,, i =0,1,2
\end{aligned}
$ & $c_{2}^{(0)}$ & $c_{2}^{(0)}=\frac{32+5d_{0}^{(4)}-d_{0}^{(5)}}{5!}$\tabularnewline
\hline 
\hline 
$a^{6}$ & $\begin{aligned}c_{j}^{(i)} & =0,\,\,i=0,1,2,\,\,j=5,9,10\\
c_{j}^{(i)} & =0,\,\,i=1,2,\,\,j=6,7,8\\
c_{8}^{(0)} & =-\frac{4}{45}-c_{6}^{(0)}\\
f_{1}^{(0)} & =0\\
d_{1}^{(0)} & =-\frac{8}{45}+c_{6}^{(0)}+4c_{7}^{(0)}
\end{aligned}
$ & $c_{6}^{(0)},c_{7}^{(0)}$ & $\begin{aligned}
c_{6}^{(0)}= & \frac{176\,+\,d_{0}^{(4+5+6)}\,+\,d_{1}^{(6)}}{180}\\
c_{7}^{(0)}= & -\frac{128\,+\,d_{0}^{(4+5+6)}}{6!}

\end{aligned}
$\tabularnewline
\hline 
\end{tabular}
\end{center}
\caption{ Constraints on the free coefficients of the up to $a^6$, gravitational Compton ansatz \eqref{eq:ansatzspin}  if  the \textit{Spin-shift-symmetry} was imposed. The last column shows the relation to  \cite{Aoude:2022trd}, where the symmetry  was imposed directly on the 2 PM amplitude.
Operators $c_{4,9,10}^{(i)}$ are proportional to $|a|$, and  are not invariant under such symmetry; this is the reason they did not show up in  the analysis of \cite{Bern:2022kto,Aoude:2022trd}. }
\label{tab:spin-shift-symmetry}
\end{table}

As it turns out, at lower orders in spin,  the \textit{spin-shift symmetry} can   be traced to be a symmetry of the  opposite-helicity gravitational Compton amplitude. The symmetry  transformation is  given by \cite{Aoude:2022thd}:
\begin{equation}\label{eq:shift_sym}
 a^\mu\to a^\mu+\varsigma_b q^\mu/q^2  \,,  
\end{equation}
 where $q=k_2^\mu+k_3^\mu$, and $\varsigma_b$ is an arbitrary parameter\footnote{The  analog symmetry transformation  $\Delta a^\mu\alpha\, k_2^\mu$ leaves invariant the  3-point amplitude \eqref{eq:classical3pt}. }. It can be  easily checked the exponential  in \eqref{eq:qftamp4clas} is indeed invariant under such transformation. However, one must notice this is a symmetry only  of the classical amplitude, namely after   the boost operation \eqref{eq:boost_4}, followed by the  merging of the two exponentiated quantum operators, are performed.  In other words, the quantum amplitude \eqref{eq:4ptexpdef} does not possess such symmetry. 
 
Given its emergence in the classical limit, one could ask if such symmetry is a feature of the BHPT analysis and can be traced back to the Teukolsky equation. To address this, we observe how such transformation acts on  the same helicity Compton amplitude. In this case, the classical amplitude  is given by an analog exponential \cite{Aoude:2020onz}
\begin{equation}
\label{eq:same_helicity_compton}
    A=\langle A_0\rangle e^{-(k_2+k_3)\cdot a}\,,
\end{equation}
 which does  not remain invariant under 
 the transformation \eqref{eq:shift_sym}. It is a short exercise to further check directly that the covariant amplitude \eqref{eq:a4CLASSSPIN2} does not possess the symmetry except in the opposite helicity case. As this general helicity setup is captured universally by the Teukolsky equation (in particular \eqref{eq:same_helicity_compton} is consistent with BHPT at least up to $a^6$) this suggests that the symmetry is not a property of the equation but rather a property of the 2PM amplitude. Indeed, as the same-helicity setup does not contribute to the 2PM amplitude, this explains why the latter enjoys shift symmetry.\footnote{This was explicitly checked in \cite{Aoude:2022trd}, up to eighth  order in spin for generic spin orientation. }

At this point, we could also inquire  how such a symmetry  constrains our ansatz \eqref{eq:ansatzspin} for   the higher-spin amplitude. Indeed, after imposing  the constraint for cancellation of   the  spurious pole (see column 2 of \cref{tab:Teukolskysolutions}),  invariance under the transformation   \eqref{eq:ansatzspin}, fixes the free coefficients as    summarized in  \cref{tab:spin-shift-symmetry}. This is in agreement with the analysis of  \cite{Aoude:2022trd}.
In this work, however,  we have kept free the coefficients of the opposite helicity Compton ansatz and instead,  asked whether the  shift-symmetry arises from the BHPT analysis. As it turns out,   solutions of the Teukolsky equation, at the given order in spin $a^n$,  do not preserve such symmetry for $n>4$.

\section{Covariant spin multipole double copy }\label{eq:spin_dc}

In this appendix, we show how to obtain the  classical gravitational  Compton amplitude \eqref{eq:a4CLASSSPIN2} up to quadratic order in spin, from the covariant spin multipole double copy introduced in \cite{Bautista:2019tdr,Bautista:2021inx}. We first proceed by  introducing  the formalism in general dimensions, and then specialize to the $D=4$ case, in order to make contact with the BHPT results. 

Let us start by recalling the spin multipole decomposition of the QED Compton amplitude for matter of spin $s-0,1/2,$ and $1$, with the kinematic conventions outlined in \cref{eq:A4-figure},  as given in \cite{Bautista:2019tdr}
\begin{equation}
A_{s,4}^{{\rm {QED}}}=\frac{e^{2}}{\left(p_{1}{\cdot}k_{2}\right)\left(p_{1}{\cdot}k_{3}\right)}\bra{\varepsilon_{s,f}}{\cdot}\left[{\omega^{(0)}}J_s^{(0)} {+}\omega_{\mu\nu}^{(1)}J_s^{(1)\mu\nu}{+}\omega_{\mu\nu\rho\sigma}^{(2)}J_s^{(2)\mu\nu\rho\sigma}  \right]{\cdot}\ket{\varepsilon_{s,i}}. \label{eq:comptom multipoles}
\end{equation}
The multipole coefficients $\omega^{(i)}$ are universal  and read explicitly
\begin{align}
\omega^{(0)} & =2p_{1}{\cdot}F_{2}{\cdot}F_{3}{\cdot}p_{1},\label{eq:w0}\\
\omega^{(1)\,\mu\nu} & =p_{1}{\cdot}F_{2}{\cdot}p_{4}F_{3}^{\mu\nu}{+}p_{1}{\cdot}F_{3}{\cdot}p_{4}F_{2}^{\mu\nu}{+}\frac{p_{1}{\cdot}(k_{2}{+}k_{3})}{2}\left[F_{2,\rho}^{\mu}F_{3}^{\rho\nu}{-}F_{2,\rho}^{\nu}F_{3}^{\rho\mu}\right]\label{eq:w1}\\
\omega^{(2)\,\mu\nu\rho\sigma} & = \frac{k_2{\cdot}k_3}{8}\Big[F_2^{\mu\nu}F_3^{\rho\sigma}+F_3^{\mu\nu}F_2^{\rho\sigma}\Big]\,,
\end{align}
with $F^{\mu\nu}_i = 2 k_i^{[\mu}\epsilon_i^{\nu]} $. The spin multipole operators are denoted by  $J_s^{(0)} = \mathbb{I}_s$,   $J_s^{(1)\mu\nu}=J_s^{\mu\nu}$ and $J_s^{(2)\mu\nu\rho\sigma} = \{J_s^{\mu\nu},J_s^{\rho\sigma}\}$, where $J_s^{\mu\nu}$  corresponds to the   Lorentz generator in the  spin-$s$ representation, acting on  the corresponding spin-s polarizations $\ket{\varepsilon_{s,i}}$. Note  that while $J_s^{(0)}$ and $J_s^{(1)\mu\nu}$ are  irreducible representations of the Lorentz group SO($D-1,1$),  the  operator $J_s^{(2)\mu\nu\rho\sigma}$ has the symmetries of the Riemann tensor and can be further decomposed into irreducible SO($D-1,1$) representations. This decomposition goes by the name of Ricci decomposition \footnote{See for instance the Wikipedia article \url{https://en.wikipedia.org/wiki/Ricci_decomposition}. }. When further decomposing  the quadratic in $J$ contribution to the Compton amplitude into irreps., it follows  
\begin{equation}
\omega_{\mu\nu\rho\sigma}^{(2)}J_s^{(2)\mu\nu\rho\sigma} =\left\{ \begin{matrix}\hat{1}_s[\omega^{(2)}]+[\omega^{(2)}]_{\mu\nu}Q_s^{\mu\nu},\quad s=1,\\
\hat{1}_s[\omega^{(2)}]+[\omega^{(2)}]_{\mu\nu\rho\sigma}\ell^{\mu\nu\rho\sigma},\,s=\frac{1}{2},
\end{matrix}\right.
\label{eq:dec}
\end{equation}
where $\ell_s^{\mu\nu\rho\sigma}=J_s^{(2)[\mu\nu\rho\sigma]}=\parbox{4pt}{\ytableausetup{mathmode,boxsize=0.2em}\ydiagram{1,1,1,1}},$
and
\begin{equation}
\hat{1}_s=\frac{J_{s,\mu\nu}J_s^{\mu\nu}}{2},\,\,Q_s^{\mu\nu}=\ytableausetup{mathmode,boxsize=0.4em}\ydiagram{2}=\{J_s^{\mu\rho},J_{s,\rho}^{\,\,\nu}\}+\frac{4}{D}\eta^{\mu\nu}\hat{1}_s\,.\label{eq:quadandl}
\end{equation}
We identify $Q_s^{\mu\nu}$ with  the traceless Ricci tensor, whereas $\hat{1}_s$ corresponds to the scalar curvature.  Notice remarkably, \eqref{eq:comptom multipoles} does not possess a Weyl contribution. This will be important when we discuss the double copy below.  In addition, for spin $1/2$ we get a totally antisymmetric contribution due to the non-commutativity nature of the Dirac gamma matrices. 

In \eqref{eq:dec}, we have further introduced the notation  $[\omega^{(2)}]$ ,  $[\omega^{(2)}]_{\mu\nu}$ and $[\omega^{(2)}]_{\mu\nu\rho\sigma}$ for the corresponding projections of the multipole coefficient $\omega^{(2)\mu\nu\rho\sigma}$ . The explicit form for the two first read
\begin{equation}
    [\omega^{(2)}] = \frac{4}{D(D-1)}\omega_{\mu\nu\rho\sigma}^{(2)} \eta^{\mu[\rho}\eta^{\sigma]\nu}=\frac{k_2{\cdot}k_3}{D(D-1)}F_{2,\mu\nu}F_3^{\mu\nu}
\end{equation}
\begin{equation}
    [\omega^{(2)}]_{\mu\nu}=\frac{k_2{\cdot}k_3}{D-2}F_{2\,(\mu|\rho}F_{3\,|\nu)}^{\rho}\,,
\end{equation}
whereas for the latter we simply have $[\omega^{(2)}]^{\mu\nu\rho\sigma}=\omega^{(2)\,[\mu\nu\rho\sigma]}$. 
We refer to the irreducible operators of  SO($D-1,1$) as the \textit{covariant spin multipole moments}. 
This then allows us to identify the covariant traceless spin \textit{quadrupole moment} $Q_s^{\mu\nu}$, existing only for spin $1$ particles in QED. This is the reason, electrons do not possess spin quadrupole moment.

\textit{Covariant $\frac{1}{2}\otimes\frac{1}{2}$ double copy}: 
Let us now compute the gravitational Compton amplitude up to  quadratic order in spin. It is given simply by the KLT double copy formula via 
\begin{equation}
\bra{\tilde{\varepsilon}_{\tilde{s},f},\varepsilon_{s,f}}A_{4}^{{\rm {\rm GR}}}(J^{s+\tilde{s}})\ket{\tilde{\varepsilon}_{\tilde{s},i},\varepsilon_{s,i}}=K_{4}\bra{\varepsilon_{s,f}}A_{4}^{{\rm QED}}\left(J^{(s)}\right)\ket{\varepsilon_{s,i}}\odot\bra{\tilde{\varepsilon}_{\tilde{s},i}}\tilde{A}_{4}^{{\rm QED}}\left(\tilde{J}^{(\tilde{s})}\right)\ket{\tilde{\varepsilon}_{\tilde{s},f}}\,,
\end{equation}
where the KLT kernel at four-points is 
\begin{equation}
K_{4}=\frac{\kappa^{2}}{8e^{4}}\frac{\left(p_{1}{\cdot}k_{2}\right)\left(p_{1}{\cdot}k_{3}\right)}{k_{2}{\cdot}k_{3}}.
\end{equation}

Taking each copy of the QED amplitude in this expression to be the spin $s=1/2$ amplitude, allows us to  define the  symmetric double copy product $\odot$ for the $SO(D-1,1)$ spin-multipoles as follows:
\begin{equation}
\begin{split}
   \mathbb{I}_{\frac{1}{2}}\odot \mathbb{I}_{\frac{1}{2}}&=\mathbb{I}_{1},\,\,\,\,\, \hat{1}_{\frac{1}{2}}\odot \mathbb{I}_{\frac{1}{2}} = \frac{1}{2} \hat{1}_{1}, \,\,\,\,\,  \mathbb{I}_{\frac{1}{2}}\odot J_{\frac{1}{2}}^{\mu\nu}=\frac{1}{2} J_{1}^{\mu\nu},\\
    J_{\frac{1}{2}}^{\mu\nu}\odot J_{\frac{1}{2}}^{\rho\sigma} &{=} \frac{1}{4}\Sigma_{1}^{\mu\nu\rho\sigma}{+}\frac{1}{D{-}2}\eta^{[\sigma[\nu}Q_{1}^{\mu]\rho]}
  \quad{+}\frac{1}{2D(D{-}1)}\eta^{\sigma[\nu}\eta^{\mu]\rho}\hat{1}_{1} 
\end{split}
\label{eq:dc_def}
\end{equation}
Notice the  double copy rule of the second line corresponds to  nothing but the Ricci decomposition of the product of two Lorentz generators into  irreducible representations of SO($D-1,1$). Spin multipoles  in the right-hand side of eq. \eqref{eq:dc_def} corresponds to  operators acting the gravitational theory, whereas the one in the left-hand side acts on their corresponding single  copy. 

Let us provide several explicit examples that follow from double copy rules  \eqref{eq:dc_def}: At spin zero, the double copy recovers the usual scalar gravitational Compton amplitude

\begin{equation}
    A_4^{(0)\text{GR}} = \frac{\kappa^{2}}{8}\frac{\omega^{(0)}\omega^{(0)}}{k_{2}{\cdot}k_{3}\left(p_{1}{\cdot}k_{2}\right)\left(p_{1}{\cdot}k_{3}\right)}
\end{equation}
Next, at linear order in spin we simply have 
\begin{equation}
   A_4^{(\frac{1}{2})\text{GR}} =   \frac{\kappa^{2}}{8}\frac{\omega^{(0)}\omega_{\mu\nu}^{(1)}\bra{\tilde{\varepsilon}_{1/2,f},\varepsilon_{1/2,f}}J_{1}^{\mu\nu}\ket{\tilde{\varepsilon}_{1/2,i},\varepsilon_{1/2,i}}}{k_{2}{\cdot}k_{3}\left(p_{1}{\cdot}k_{1}\right)\left(p_{1}{\cdot}k_{2}\right)}\,,
\end{equation}
and finally at quadratic order in spin 

\begin{equation}
\begin{split}
     A_4^{(1)\text{GR}} = & \frac{\kappa^{2}}{8}\frac{\frac{1}{2}\omega^{(0)}[\omega^{(2)}]\bra{\tilde{\varepsilon}_{1,f},\varepsilon_{1,f}}\hat{1}_1 \ket{\tilde{\varepsilon}_{1,i},\varepsilon_{1,i}} }{k_{2}{\cdot}k_{3}\left(p_{1}{\cdot}k_{2}\right)\left(p_{1}{\cdot}k_{3}\right)}\\
     &
     +\frac{\kappa^{2}}{8}\frac{\omega_{\mu\nu}^{(1)}\omega_{\rho\sigma}^{(1)}}{k_{2}{\cdot}k_{3}\left(p_{1}{\cdot}k_{2}\right)\left(p_{1}{\cdot}k_{3}\right)}\bra{\tilde{\varepsilon}_{1,f},\varepsilon_{1,f}} \Big[\frac{1}{4}\Sigma_{1}^{\mu\nu\rho\sigma}{+}\frac{1}{D{-}2}\eta^{[\sigma[\nu}Q_{1}^{\mu]\rho]}\\
     &\qquad\qquad\qquad\qquad\qquad\qquad
     {+}\frac{1}{2D(D{-}1)}\eta^{\sigma[\nu}\eta^{\mu]\rho}\hat{1}_1\Big]\ket{\tilde{\varepsilon}_{1,i},\varepsilon_{1,i}}\,.
\end{split}
\end{equation}
Here we have omitted the contribution form $\parbox{4pt}{\ytableausetup{mathmode,boxsize=0.2em}\ydiagram{1,1,1,1}}$, since it does not contribute to the classical amplitude. 

\textit{Classical limit}: 
As explained in Appendix A of \cite{Bautista:2019evw}, in order to interpret the results for the previously computed amplitude as those for the scattering of a gravitational wave off a spinning BHs, we need to choose a reference frame -- which can be fixed by choosing a time-like vector $u^\mu $ satisfying the Spin Supplementary Condition (SSC) -- so that the massive polarization states are aligned towards the same  canonical polarization states. 
 When doing so, the $SO(D-1,1)$ generator $J_1^{\mu\nu}$, which consists of a  $SO(D-1)$  Wigner rotation plus a boost, $J_1^{\mu\nu} =S^{\mu\nu}-2u^{[\mu}K^{\nu]}$, can be interpreted as a classical spin tensor for the rotating object, once the  boost component is removed away. The SSC  to satisfy is then simply given by  $u_\mu S^{\mu\nu}=0$.  After this is done, the  polarization states can be removed from the gravitational amplitude, leaving us with a classical object, which we will interpret as the classical amplitude for the gravitational wave scattering process.  The  alignment  of the polarization states of the incoming/outgoing massive particle effectively  induces a map of the  $SO(D-1,1)$ multipoles  towards the  $SO(D-1)$ multipoles. We refer  to the multipole moments of the rotation subgroup as \textit{rotation multipole moments}.
 For a detailed explanation of how to do the map,  the reader can consult the  aforementioned  Appendix. In here we  restrict ourselves to simply summarize the   map as follows:
\begin{equation}\label{eq:branching-general-d}
    \begin{split}
        J_1^{\mu\nu}&\rightarrow S^{\mu\nu}\\
        Q_1^{\mu\nu}&\rightarrow \bar{Q}^{\mu\nu}+ \frac{1}{D-1}\bar{\eta}^{\mu\nu} \hat{1}_1 - \frac{1}{D}\hat{1}_1 \\
        \Sigma^{\mu\nu\rho\sigma}&\rightarrow\frac{4}{D-3}u^\nu \bar{Q}^{\mu\rho}u^{\sigma}\\
        \hat{1}_1&\to \frac{1}{2}S_{\mu\nu}S^{\mu\nu},
    \end{split}
\end{equation}
where $\bar{\eta}^{\mu\nu}=\eta^{\mu\nu}-u^{\mu}u^{\nu}$, and $\bar{Q}^{\mu\nu}$ is the $SO(D-1)$ spin quadrupole moment, which satisfies the traceless condition $\bar{Q}^{\mu\nu}\bar{\eta}_{\mu\nu}=0$.

\textit{Spin-multipoles for D=4}. 
Now, since we are interested in making contact with the BHPT computations,  we specify the spin multipoles for the $D=4$ scenario.  In such case, we can write the \textit{rotation spin dipole moment} in  terms of the  Pauli-Lubanski vector $s^\mu = Ma^\mu$, via $S^{\mu\nu}=\epsilon^{\mu\nu\rho\sigma}p_{1\rho }a_{\sigma}$, whereas the \textit{rotation spin quadrupole moment}   reads
\begin{equation}\label{eq:branching-4-d}
\begin{split}
    \bar{Q}^{\mu\nu}&=m^2 \left(a^\mu a^\nu-\frac{1}{3}\bar{\eta}^{\mu\nu}a^2\right)\,.
    \end{split}
\end{equation}
In this notation, the  SSC  is satisfied by the  spin vector, $p_{1\mu} a^\mu=0$. Finally, to extract the classical limit of the double copy amplitude we have to do the usual $\hbar$ scaling of the massless momenta, $k_i\to \hbar k_1$, and in an analogous way for the spin vector we do  $a^\mu\to a^\mu/\hbar$. Notice we have  identified  $u_\mu$  with the incoming massive object's four-velocity; in principle  one could had identified $u^\mu$ with either $p_1^\mu$ or $p_4^\mu$, or the average $(p_1+p_4)^\mu/2$; however, in the classical limit all of the choices are equivalent to each other, as we widely explained in the paragraph above  \eqref{spinvec}. With all this in mind,  one can explicitly check that the final classical amplitude up to quadratic order in spin can be written as:
\begin{equation}
    \langle A_4 ^{\text{GR}}\rangle = \frac{\kappa^2}{8}\frac{\langle\omega^{(0)}\rangle}{k_2{\cdot}k_3 (p_1{\cdot}k_2)^2}\left[\langle \omega^{(0)} \rangle + \langle \omega^{(1)\mu\nu}\rangle\epsilon_{\mu\nu\rho\sigma}p_1^{\rho}a^{\sigma} + \langle \omega^{(2)}_{\alpha\beta} \rangle a^{\alpha}a^{\beta}   \right]+ \mathcal{O}(a^3)\,,\label{eq:a4CLASSSPIN23}
\end{equation}
which corresponds exactly to \eqref{eq:a4CLASSSPIN2}. In here the angular brackets indicate the classical limit of the corresponding multipole  coefficients given in \eqref{eq:w0} and \eqref{eq:w1}. We have also identify the classical multipole coefficient for the quadratic in spin contribution in  classical E\&M as 
\begin{align}
  \langle \omega^{(2)\alpha\beta} \rangle &=\Big[p_{1}{\cdot}F_{2}{\cdot}F_{3}{\cdot}p_{1}q_{\mu}\mathcal{P}^{\mu\nu\alpha\beta}q_{\nu}{+}2k_{2}{\cdot}k_{3}m^{2}\Big(\mathcal{P}^{\mu\nu\alpha\beta}{+}\frac{\eta^{\mu\nu}\eta^{\alpha\beta}}{2}\Big)F_{2}^{(\mu|\delta}F_{3}^{\gamma|\nu)}\eta_{\gamma\delta}\Big]\,,\label{eq:quad_class}\\
   \mathcal{P}^{\mu\nu\alpha\beta}&=\frac{\eta^{\mu\alpha}\eta^{\nu\beta}+\eta^{\mu\beta}\eta^{\nu\alpha}}{2}-\eta^{\mu\nu}\eta^{\alpha\beta}\,,
\end{align}
where $q^{\mu}=k_3^\mu-k_2^\mu$.
Notice remarkably that after combining  the different contributions to the  \textit{rotation spin multiple moments},  the Classical GR amplitude has the factorization form $A^{\text{GR}} = \langle A_0^{\text{QED}}\rangle \times \langle A_s^{\text{QED}}\rangle $. The quadratic in spin contribution  was originally  given in the ancillary files of \cite{Bautista:2021inx}, whose unitarity gluing with the 3-point amplitude, recovers the quadratic in spin  two-body radiation amplitude obtained in \cite{Jakobsen:2021lvp}.

Let us finalize this appendix commenting on the \textit{spin-shift-symmetry} of \eqref{eq:shift_sym}. For arbitrary helicity configurations of the massless legs, \eqref{eq:a4CLASSSPIN23} does not possess such a symmetry. We can however  show for the opposite helicity configuration, such symmetry is manifested when starting from the amplitude written in vector notation. For that, let us recall for this configuration we can choose the  gauge \eqref{eq:chkgauge}, in which the polarization vectors of the massless particles are proportional, say $\epsilon_2^+=\epsilon_3^+=\epsilon^+$. Using this into \eqref{eq:a4CLASSSPIN23}, $\langle A_4^{\text{GR}}\rangle^{++}$ becomes
\begin{equation}\label{eq:com}
    \langle A_4 ^{\text{GR}}\rangle^{++} = \frac{\kappa^2}{8}\frac{\langle\omega^{(0)}\rangle^{++}}{k_2{\cdot}k_3 (p_1{\cdot}k_2)^2}\left[\langle \omega^{(0)} \rangle^{++}+\langle\omega^{(1)\mu\nu}\rangle^{++} \epsilon_{\mu\nu\rho\sigma}p_1^{\rho}a^{\sigma} + \langle \omega^{(2)}_{\alpha\beta} \rangle ^{++}a^{\alpha}a^{\beta}   \right]+ \mathcal{O}(a^3)\,,
\end{equation}
where the  multipole coefficients are  $\langle\omega^{(0)}\rangle^{++}= -2 k_2{\cdot}k_3 (p_1{\cdot}\epsilon^+)^2$ for the scalar part,   the dipole piece is  $\langle\omega^{(1)\mu\nu}\rangle^{++}= k_2{\cdot}k_3 p_1{\cdot}\epsilon^+ F_q^{+\,\mu\nu} $, with $F_q^{+\,\mu\nu}= 2 \epsilon^{+\,[\mu}q^{\nu]} $, and finally the quadrupole coefficients is $\langle \omega^{(2)}_{\alpha\beta} \rangle ^{++}a^\alpha a^\beta=k_2{\cdot}k_3\left[\left((q{\cdot}a)^2-q^2 a^2\right)(p_1{\cdot}\epsilon^+)^2 -2m^2 (a{\cdot}\epsilon^+)^2\right]$. 
The dipole coefficient is manifestly invariant under the transformation \eqref{eq:shift_sym} since it is proportional to $ \epsilon_{\mu\nu\rho\sigma}q^\mu  a^\nu$.   For the quadrupole coefficient, one can explicitly check that after using the identity \eqref{eq:a2identity} for the quadratic Casimir\footnote{Recall here we have to change the sign of $k_3$ to follow the conventions of GW scattering introduced in \eqref{eq:sandtclassical}. }, the quadrupole coefficient is left invariant under the transformation \eqref{eq:shift_sym}, therefore we conclude \eqref{eq:com} is \textit{spin-shift-symmetric}.

\section{Plane waves in Kerr space-time}
\label{PW-App}

In this appendix, we will review the construction of plane wave solutions on a Kerr background.

\subsection{Polar scattering}

We first focus on the simpler case of  the scattering of a plane wave moving parallel to the axis of rotation of the BH. Since we work in Boyer-Lindquist coordinates this is a plane wave moving up the z-axis. This construction follows closely the descriptions in sections III and IV of \cite{Chrzanowski:1976jb}. The strategy is to first work in flat spacetime, as a representation of asymptotic infinity in the black hole spacetime. After constructing the harmonic modes of $\psi_4$ for a flat space plane wave, we then make the replacement $r\rightarrow r_*$, to account for the long range nature of the black hole potential, see e.g. \cite{Matzner1968} .

In flat spacetime, the metric perturbation for a plane wave moving up the z-axis is given by
\begin{align}
	h_{\mu\nu}=H\Re\Big(\varepsilon_\mu\varepsilon_\nu e^{ik\cdot x}\Big)=H\begin{pmatrix}
0 & 0 & 0 & 0\\
0 & \cos(\omega\chi_0) &  \sin(\omega\chi_0) & 0\\
0 &  \sin(\omega\chi_0) & - \cos(\omega\chi_0) & 0\\
0 & 0 & 0 & 0
\end{pmatrix} 
\label{Eq:zaxisPW},
\end{align}
where $\chi_0=t-z$  and $k^\mu=(\omega,0,0,\omega)$, $\varepsilon^\mu=(0,1,-i,0)$. With this, we can construct the perturbed Riemann tensor.  Projecting then onto the flat spacetime limit of the Kinnersley tetrad
\begin{align}
l^\alpha&=\frac{1}{\Delta}(r^2+a^2,\Delta,0,a), \\
n^\alpha&=\frac{1}{2 \Sigma}(r^2+a^2,-\Delta,0,a), \\
m^\alpha&=-\frac{\bar{\varrho}}{\sqrt{2}}(i a \sin \theta,0,1,\frac{i}{\sin \theta}), \\
\bar{m}^\alpha&=-\frac{\varrho}{\sqrt{2}}(-i a \sin \theta,0,1,\frac{-i}{\sin \theta}),
\end{align}
the Weyl scalars
\begin{align}
\psi_0&=-C_{\alpha\beta\gamma\delta}l^\alpha m^\beta l^\gamma m^\delta, \\
\psi_4&=-C_{\alpha\beta\gamma\delta}n^\alpha \bar{m}^\beta n^\gamma \bar{m}^\delta,
\end{align}
are readily computed to be
\begin{align}
	\psi_0^{\rm PW}&=-\frac{1}{2}H \omega^2(1-\cos\theta)^2e^{-i\omega(t-z)}e^{2 i\phi},\\
	\psi_4^{\rm PW}&=-\frac{1}{8}H \omega^2(1+\cos\theta)^2e^{-i\omega(t-z)}e^{2 i\phi}.
\end{align}
We now wish to project onto the spin-weighted spheroidal harmonics, to obtain the modes. Writing $z=r\cos\theta$, the angular integrals can be evaluated using the stationary phase approximation (see for instance \cite{Bautista:2021wfy}). We arrive at the leading order behavior
\begin{align}
	\psi_0^{\rm PW}&\approx -4i\pi H\omega\sum_{l=2}^{\infty}\frac{1}{r}e^{-i\omega(r+t)}\sSlm{2}{\ell}{2}(\pi,0,a \omega)\sSlm{2}{\ell}{2}(\theta,\phi,a \omega), \label{Eq:psi0LO}\\
	\psi_4^{\rm PW}&\approx i\pi H\omega\sum_{l=2}^{\infty}\frac{1}{r}e^{i\omega(r-t)}\sSlm{-2}{\ell}{2}(0,0,a \omega)\sSlm{-2}{\ell}{2}(\theta,\phi,a \omega) .\label{Eq:psi4LO}
\end{align}

Obtaining the subdominant ($r^{-5}$) terms here is somewhat subtle. Since we have been using a flat space approximation, constructing higher order terms from the stationary phase approximation to the integrals gives incorrect asymptotic behavior. 

However, it turns out that the flat spacetime approximation is nonetheless sufficient to obtain the subdominant pieces by making use of the Teukolsky-Starobinsky (TS) identities. The TS identities are most succinctly written using the Geroch-Held-Penrose (GHP) notation as (see e.g. \cite{Pound:2021qin}):
\begin{align}
  \th^4\zeta^{4} \psi_4
    &= \edth'^4 \zeta^{4} \psi_0 - 3 M\mathcal{L}_t \bar{\psi}_0,\label{Eq:TS1}\\
    \th'^4 \zeta^{4} \psi_0 &= \edth^4 \zeta^{4} \psi_4 + 3 M\mathcal{L}_t  \bar{\psi}_4.\label{Eq:TS2}
\end{align}
Here, $\zeta=(r-i a \cos\theta)$, $\mathcal{L}_t $ is the Lie derivative along the timelike killing vector, and the differential operators $\th$, said `thorn', and $\edth$, said `edth', and their primes are given by
\begin{align}
    \th \chi&=(l^\mu\partial_\mu-p \epsilon-q\bar{\epsilon})\chi,\\
    \th' \chi&=(n^\mu\partial_\mu+p \epsilon'+q\bar{\epsilon}')\chi, \\
    \edth \chi&=(m^\mu\partial_\mu-p \beta+q\bar{\beta}')\chi,\\
    \edth \chi&=(\bar{m}^\mu\partial_\mu+p \beta'-q\bar{\beta})\chi.
\end{align}

The set of integers $\{p,q\}$  is the GHP weights of the function $\chi$ being acted upon. In the Kinnersley tetrad $\epsilon=\epsilon'=0$, $\beta=\frac{\cot\theta}{2\sqrt{2}(r+ia \cos\theta)}$, $\beta'=\frac{r \cot \theta -i a \sin \theta  \left(\csc ^2\theta +1\right)}{2\sqrt{2}(r-ia\cos\theta)}$. The weightings of the relevant quantities are
\begin{align}
    \th&:\{1,1\}  ,\\
    \edth&:\{1,-1\} , \\
    \psi_4&:\{-4,0\} ,\\
    \psi_0&:\{4,0\} ,\\
    \zeta&:\{0,0\} .
\end{align}
and $\bar{\chi}$ has weight $\{q,p\}$, whereas $\chi'$ has $\{-p,-q\}$.  For  $\psi_4$ we find that 
\begin{align}
    \th^4\zeta^{4}\left[r^{-5}e^{-i\omega(t+r)}\sSlm{-2}lm(\theta,\phi)\right]&=16\omega^4 r^{-1}e^{-i\omega(t+r)}\sSlm{-2}lm(\theta,\phi)+\mathcal{O}(r^{-2}),
\end{align}
and for $\psi_0$ that 
\begin{align}
     \edth'^4\zeta^{4}\left[r^{-1}e^{-i\omega(t+r)}\sSlm{2}lm(\theta,\phi)\right]&=\frac{1}{4}\mathcal{C}_{lm}r^{-1}e^{-i\omega(t+r)}\sSlm{-2}lm(\theta,\phi)
\end{align}
where the TS constant is
\begin{equation}\label{eq:tsconstant}
    \begin{split}
        \mathcal{C}_{lm}^2=&[(\lambda+2)^2+4 a m \omega-4a^2\omega^2][\lambda^2+36 a m \omega-36 a^2\omega^2]\\
        &+(2\lambda+3)(96 a ^2\omega^2-48 a m \omega)-144 a^2 \omega^2\,,
    \end{split}
\end{equation}
where $\lambda={}_{-2}\lambda_{lm}$ is the $s=-2$ spheroidal eigenvalue.
Using \eqref{Eq:psi0LO} in \eqref{Eq:TS1} determines the subleading term for $\psi_4$, giving the asymptotic form 
\begin{align}
    \psi_4^{\rm PW}\approx H \sum_{l=2}^{\infty}\Bigg[&\frac{i\pi \omega}{r}e^{i\omega(r-t)}\sSlm{2}{\ell}{2}(0,0,a \omega)\sSlm{-2}{\ell}{2}(\theta,\phi,a \omega)+\mathcal{A}_4^+\frac{1}{r^5}e^{i\omega(r+t)}\sSlm{-2}{\ell}{2}(\theta,\phi,a \omega)\nonumber\\
    &+\mathcal{A}_4^-\frac{1}{r^5}e^{-i\omega(r+t)}\sSlm{-2}{\ell}{-2}(\theta,\phi,-a \omega)\Bigg],
\end{align}
where
\begin{align}
    \mathcal{A}_4^+&=-\frac{i}{32 \omega^3}\mathcal{C}_{\ell m \omega}\,\sSlm{-2}{\ell}{2}(0,0,a \omega),\\ %\sSlm{2}{\ell}{2}(\pi,0,a \omega) \\
    \mathcal{A}_4^-&=\frac{3}{8\omega^2}\,\sSlm{-2}{\ell}{2}(0,0,a \omega).
\end{align}

\subsection{Off axis scattering}

We now consider a plane wave approaching the Kerr black hole from an arbitrary angle $\gamma$ off the axis of rotation. For simplicity, we assume it moves in a plane with the $x$-axis as a normal when $\gamma=\pi$.

We begin by rotating the flat space $z$-axis plane wave by an angle $\gamma $ about the $y$-axis. This gives the metric perturbation
% \begin{align}
% h_{\mu\nu}=H
%     \left(
% \begin{array}{cccc}
%  0 & 0 & 0 & 0 \\
%  0 & \cos ^2(\gamma ) \cos ( \omega \chi  ) & \cos (\gamma ) \sin ( \omega \chi  ) & \sin (\gamma ) \cos (\gamma ) \cos (\chi 
%   \omega ) \\
%  0 & \cos (\gamma ) \sin ( \omega \chi  ) & -\cos ( \omega \chi  ) & \sin (\gamma ) \sin ( \omega \chi  ) \\
%  0 & \sin (\gamma ) \cos (\gamma ) \cos ( \omega \chi  ) & \sin (\gamma ) \sin ( \omega \chi  ) & \sin ^2(\gamma ) \cos (\chi 
%   \omega ) \\
% \end{array}
% \right)
% \end{align}
\begin{align}
    h_{\mu\nu}=H
    \left(
\begin{array}{cccc}
 0 & 0 & 0 & 0 \\
 0 & \cos ^2(\gamma ) \cos (\chi  \omega ) & \cos (\gamma ) \sin (\chi  \omega ) & -\sin (\gamma ) \cos (\gamma ) \cos
   (\chi  \omega ) \\
 0 & \cos (\gamma ) \sin (\chi  \omega ) & -\cos (\chi  \omega ) & -\sin (\gamma ) \sin (\chi  \omega ) \\
 0 & -\sin (\gamma ) \cos (\gamma ) \cos (\chi  \omega ) & -\sin (\gamma ) \sin (\chi  \omega ) & \sin ^2(\gamma ) \cos
   (\chi  \omega ) \\
\end{array}
\right)
\end{align}
where after the active rotation $\chi_0$ is replaced by $\chi=t-r \sin (\gamma ) \sin (\theta ) \cos (\phi )-r \cos (\gamma ) \cos (\theta )$.

The Weyl scalars are then
\begin{align}
    \psi_0^{\rm PW}&=2 H e^{-i\omega\chi-2i\phi}\omega^2\left(\sin \left(\frac{\gamma }{2}\right) \cos \left(\frac{\theta }{2}\right)- e^{i \phi } \cos \left(\frac{\gamma
   }{2}\right) \sin \left(\frac{\theta }{2}\right)\right)^4,\\
    \psi_4^{\rm PW}&=\frac 1 2 H e^{-i\omega\chi-2i\phi}\omega^2\left(e^{i \phi } \cos \left(\frac{\gamma }{2}\right) \cos \left(\frac{\theta }{2}\right)+ \sin \left(\frac{\gamma
   }{2}\right) \sin \left(\frac{\theta }{2}\right)\right)^4.
\end{align}
Projecting these onto spheroidal harmonics is more complicated in this situation. Each of the Weyl scalars take the general form
\begin{align}
    \psi=e^{-i(\omega t-\beta)}\sum_{k=0}^4  A_k(\theta,\gamma)e^{i(\alpha\cos\phi+(k-2)\phi)}
\end{align}
with $\alpha=\omega r \sin\gamma \sin\theta,\beta=\omega r\cos\gamma\cos\theta$. We will need to compute integrals of the form
\begin{align}
    \psi_{lm}=\int_{-1}^1e^{-i(\omega t-\beta)}\sum_{k=0}^4A_k\left[\int_0^{2\pi}e^{i(\alpha\cos\phi+(k-2-m)\phi)}d\phi\right]\sSlm slm(x,0)dx
\end{align}
with $x=\cos \theta$. Using the identity
\begin{align}
    J_\nu(\alpha)=\frac{1}{2\pi i^\nu}\int_0^{2\pi}e^{i(\alpha\cos\phi-\nu\phi)}d\phi
\end{align}
where $ J_\nu(\alpha)$ is the Bessel function of the 1st kind, the integral in the square brackets is immediate. Employing the asymptotic form
\begin{align}
    J_\nu(\alpha)\sim\frac{1}{\sqrt{2\pi\alpha}}\left(-e^{i(\alpha+\frac{3\pi}{4}-\frac{\pi\nu}{2})}+e^{-i(\alpha-\frac{\pi}{4}-\frac{\pi\nu}{2})}\right)
\end{align}
as $|\alpha|\rightarrow\infty$, the remaining integrals are
\begin{align}
    \psi_{lm}=e^{-i\omega t}\sqrt{\frac{2\pi}{\alpha}}\left[\sum_{k=0}^4A_k'\int_{-1}^1e^{i(\alpha+\beta)}\sSlm slm(x,0)dx+\sum_{k=0}^4A_k''\int_{-1}^1e^{i(\beta-\alpha)}\sSlm slm(x,0)dx\right].
\end{align}
where we have made the redefinitions $A_k'=-e^{\frac{3i\pi}{4}}A_k$ and $A_k''=e^{i(\frac{\pi}{4}+\pi\nu)}A_k$. It is east to see that $\alpha+\beta=\omega r \cos(\theta-\gamma)$ and $\beta-\alpha=\omega r \cos(\theta+\gamma)$. Thus we need to compute the integrals
\begin{align}
    I_1=\int_0^\pi e^{i\omega r \cos(\theta-\gamma)}f(\theta)\sin\theta d\theta, \\
    I_2=\int_0^\pi e^{i\omega r \cos(\theta+\gamma)}f(\theta)\sin\theta d\theta.
\end{align}
Using a stationary phase approximation these are
\begin{align}
    I_1&=\sqrt{\frac{2\pi}{\omega r}}e^{-i\pi/4}f(\gamma)\sin\gamma~  e^{i\omega r}, \\
    I_2&=\sqrt{\frac{2\pi}{\omega r}}e^{i\pi/4}f(\pi-\gamma)\sin\gamma~ e^{-i\omega r}.
\end{align}
Using the explicit results for the Weyl scalars we find
\begin{align}
    \psi^{\rm PW}_{0,lm}&=H\frac{4 i\pi \omega}{r}(-1)^m\sSlm{2}lm(\pi-\gamma,0,a\omega)e^{-i\omega(t+r)},\\
    \psi^{\rm PW}_{4,lm}&=-H\frac{ i\pi \omega}{r}\sSlm{-2}lm(\gamma,0,a\omega)e^{-i\omega(t-r)}.
\end{align}
Once again, upon using the Teukolsky-Starobinsky identities we find
\begin{align}
    \psi_4^{\rm PW}\approx H \sum_{l=2}^{\infty}\sum_{m=-l}^{l}\Bigg[&\frac{i\pi \omega}{r}e^{i\omega(r-t)}\sSlm{2}{\ell}{m}(\gamma,0,a \omega)\sSlm{-2}{\ell}{m}(\theta,\phi,a \omega)\nonumber\\
    &+\mathcal{A}_{4,lm}^{+}\frac{1}{r^5}e^{i\omega(r+t)}\sSlm{-2}{\ell}{m}(\theta,\phi,a \omega)+\mathcal{A}_{4,lm}^{-}\frac{1}{r^5}e^{-i\omega(r+t)}\sSlm{-2}{\ell}{m}(\theta,\phi,-a \omega)\Bigg],
\end{align}
where
\begin{align}
    \mathcal{A}_{4,lm}^+&=\frac{i\pi}{16 \omega^3}(-1)^{l+m}\mathcal{C}_{\ell m \omega}\,\sSlm{-2}{\ell}{m}(\gamma,0,a \omega),\\ %\sSlm{2}{\ell}{2}(\pi,0,a \omega) \\
    \mathcal{A}_{4,lm}^-&=\frac{3\pi}{4\omega^2}(-1)^{m}\sSlm{-2}{\ell}{m}(\pi-\gamma,0,-a \omega).
\end{align}

\bibliographystyle{JHEP}
\bibliography{references}

\end{document}